\documentclass{emulateapj}

\usepackage{natbib}
\usepackage{amsmath}

\shorttitle{The nature of GC-IRS13E}
\shortauthors{Fritz et al.}

\begin{document}
\title{GC-IRS13E - A puzzling association of three early-type stars}
\author{ T.K.~Fritz\altaffilmark{1}, S.~Gillessen\altaffilmark{1}, K.~Dodds-Eden\altaffilmark{1},  F.~Martins\altaffilmark{2}, H.~Bartko\altaffilmark{1},   
R.~Genzel\altaffilmark{1,3},
T.~Paumard\altaffilmark{4},
T.~Ott\altaffilmark{1},
O.~Pfuhl\altaffilmark{1},
S.~Trippe\altaffilmark{5},
F.~Eisenhauer\altaffilmark{1},
D.~Gratadour\altaffilmark{4},
}

\altaffiltext{1}{Max Planck Institut f{\"u}r Extraterrestrische Physik, Postfach 1312, D-85741, Garching, Germany; tfritz@mpe.mpg.de}
\altaffiltext{2}{GRAAL-CNRS, Universit{\'e} Montpellier II -UMR5024, Place Eugme Bataillon, F-34095, Montpellier, France}
\altaffiltext{3}{Department of Physics, University of California, Berkeley, 366 Le Comte Hall, Berkeley, CA 94720-7300}
\altaffiltext{4}{LESIA, Observatoire de Paris, CNRS, UPMC, Universit{\'e} Paris Diderot; 5 Place Jules Janssen, 92190 Meudon, France}
\altaffiltext{5}{Institute de radioastronomie Millim{\'e}trique, 300 rue de la Piscine, F-38406 Grenoble, France}


\begin{abstract}
We present a detailed analysis of high resolution near-infrared imaging and spectroscopy of the potential star cluster IRS13E very close to the massive black hole in the Galactic Center. 
We detect 19 objects in IRS13E from Ks-band images, 15 of which are also detected reliably in H-band. We derive consistent proper motions for these objects from the two bands. Most objects share a similar westward proper motion. 
We characterize the objects using spectroscopy (1.45 to $2.45\,\mu$m) and (narrow-band) imaging from H- ($1.66\,\mu$m) to L'-band ($3.80\,\mu$m).
Nine of the objects detected in both Ks- and H-band are very red, and we find that they are all consistent with being warm dust clumps.
The dust emission may be caused by the colliding winds of the two Wolf-Rayet stars in the cluster.
Three of the six detected stars do not share the motion or spectral properties of the three bright stars.
This leaves only the three bright, early-type stars as potential cluster members. It is unlikely that these stars are a 
chance configuration. Assuming the presence of an IMBH, a mass of about $14000\,M_{\odot}$ follows from the velocities and positions of these three stars. However, our acceleration limits make such an IMBH nearly as unlikely as a chance occurrence of such a star association. Furthermore, there is no variable X-ray source in IRS13E despite the high density of dust and gas.
Therefore, we conclude that is unlikely that IRS13E hosts a black hole massive enough to bind the three stars. 
\end{abstract}

\maketitle

\section{Introduction}

The innermost parsec of the Galaxy hosts the supermassive black hole (SMBH) Sgr~A* \citep {Schoedel_02, Ghez_03}, accompanied by a population of young WR/O-stars \citep {Forrest_87, Allen_90, Krabbe_91,Genzel_96}. Most of these stars reside in one or two disk-like structures \citep{Genzel_03, Paumard_06, Lu_09, Bartko_09}.
A group of at least three such bright stars called IRS13E at a distance of $3.5'' =0.13\,$pc from Sgr~A* is of special interest. This group has a diameter of about $0.5''$. \citet {Maillard_04} 
identified two of the stars, E2 and E4, from their emission lines as early-type stars. In addition they identified four other early-type stars in IRS13E from broad-band SED fitting. 
The four brightest stars share a common proper motion \citep {Ott_02}.
\citet {Maillard_04} concluded without using a statistical test that such an association of young stars cannot be a coincidence.
From the radial velocities of two of the stars they estimated a cluster mass of at least $750\,M_{\odot}$ if the two stars are bound.
This mass is higher than the stellar mass seen in IRS13E. They explained the additional mass by the presence of an intermediate mass black hole (IMBH). 

In the simulations of stellar clusters of \citet{Zwart_02} the core of a dense cluster collapses and forms an IMBH. Such an IMBH would be necessary, if IRS13E were an inspiraling cluster that survives the infall into the Galactic Center (GC) and reaches the central parsec before disintegrating \citep{Hansen_03}. A cluster without central mass would be disrupted \citep{Gerhard_01} by the tidal forces of the SMBH.

\citet{Schoedel_05} measured the proper motions of the four brightest sources more accurately and estimated that the cluster has a mass of about $10000$ to $ 50000\,M_{\odot}$ if it is gravitationally bound. According to these authors, an IMBH of this mass is unlikely, mainly because of the lack of radio and X-ray emission in IRS13E, despite the presence of a lot of dust. They suggested that IRS13E could also be a cluster in the process of dissolution or a chance association. 

\citet{Paumard_06} identified the spectral types of the three brightest early-type stars. In addition, they measured the stellar surface density around IRS13E on a deconvolved H-band image. 
Inside a radius of $0.30''$, the core of IRS13E, they found at least twelve stars. Furthermore, they found an over-density out to $0.68''$ which has a total significance of $4.5\,\sigma$. They concluded that IRS13E is a cluster. Given that velocities had been measured for only four of the stars, \citet{Paumard_06} argued that the total velocity dispersion could be small. Then no dark mass is needed to explain IRS13E.
\citet{Trippe_08} measured the velocities of the stars between $0.30''$ and $0.68''$. The velocities of most of these stars are different from the stars in the center. Therefore most of them are not cluster members and thus the overdensity is less significant compared to \citet{Paumard_06}.

IRS13E2 and IRS13E4 are candidate members of the face-on counterclockwise disk \citep{Paumard_06,Bartko_09}. In this case the physical distance of these stars to Sgr~A* is identical with the projected one. According to \citet{Paumard_04, Paumard_06} IRS13E is embedded in the bar of the minispiral at a distance of z$=7-20$'' of Sgr~A* \citep{Liszt_03}. In this case IRS13E is rather far away from Sgr~A* and IRS13E2 and IRS13E4 would not be candidates members of the counterclockwise disc \citep{Bartko_09}.

Accordingly, the nature of IRS13E is still a matter of debate. This is largely because the proper motions and the nature of the member objects are known only for very few sources in the core with a radius of about $0.3''$. In particular, it is not clear whether all objects identified by previous works actually are stars. In this paper we analyze seven years of imaging data (presented in section 2).
We identify 19 objects in the core of IRS13E, for all of which we can measure proper motions (section 3). For the brightest sources E1 and E2, we obtain acceleration limits which start to to constrain the nature of IRS13E (section 3). From H- to L'-band photometry and H+K-band spectra we constrain the nature of the objects (section 4).
In Section 5 we calculate the probability that our acceleration limits occur for the two cases that a) IRS13E is bound by an IMBH and b) that it is a chance association.
Section 6 considers other data in a more qualitative fashion and theoretical models. 
Finally we summarize and conclude in Section 7. 
We assume a distance to the GC of $R_0=8\,$kpc \citep{Reid_93} and a mass of the SMBH of $M=4\times 10^6 M_{\odot}$ \citep{Ghez_08,Gillessen_09}.

\section{Dataset}
This study uses seven years of adaptive optics based imaging data and spectroscopic data obtained with an integral field spectrograph in 2004 and 2009. This section briefly describes our data set.

\subsection{NACO}\label{naco}

We obtained images with NAOS/CONICA (NACO) mounted on UT4 at the VLT \citep{Lenzen_etal2003_NACO, Rousset_etal_2003_NACO}\footnote{We use NACO data from ESO programs 70.B-0649, 71.B-0077, 71.B-0078, 71.B-0365, 072.B-0285, 073.B-0084, 073.B-0085, 073.B-0665, 073.B-0775, 075.B-0093, 077.B-0014, 078.B-0136, 179.B-0261, 179.B-0932, 183.B-0100.}. 
The camera CONICA, together with the adaptive optics (AO) system NAOS, achieves diffraction limited images in the near-infrared. 
For this study, we use mainly the $13\,$mas/pixel image scale. On this scale, the point spread function (PSF) is oversampled. This is suited better
for highly crowded fields than a strict Nyquist sampling \citep{Trippe_09}. 

For source detection and astrometry, we use data from nearly every observing run from April 2002 to May 2009, excluding only the worst 10\% of the observing runs by eye. 
In total, our sample consists of 37 H-, 74 Ks- and 5 L'-band images. 

We base the broad-band photometry on four H-, four Ks- and three L'-band images that were of particularly good quality. Furthermore, we use four narrow-band images for photometry, which we obtained at $2.06\,\mu$m, $2.24\,\mu$m (12-06-2004), $2.17\,\mu$m and $2.33\,\mu$m (13-06-2004).

All images are flat-fielded, bad pixel corrected and sky subtracted. In addition, we correct images in which saturation is important for the nonlinearity of the detector \citep{Fritz_09}.

We combine single images of typically 30 seconds exposure into stacked images. Because the faint additional sources in IRS13E lie in the PSF halos of the bright stars, the prime goal of the data reduction is to obtain the highest Strehl ratio possible for each data set, even at the cost of smaller total exposure \citep{Fritz_09}. For this reason, we select  frames of good quality (small FWHM) for a combined image of an observing run only.

\begin{figure}
\begin{center}
\includegraphics[width=0.946\columnwidth]{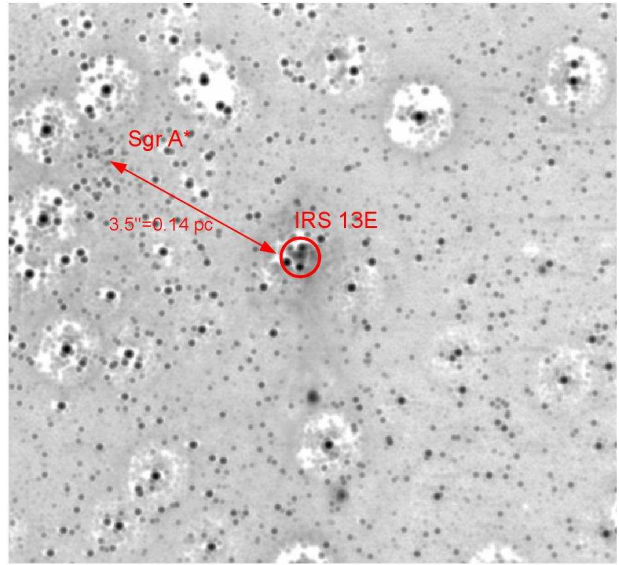}
\caption{Lucy-deconvolved Ks-image of the GC from 13-03-2008, smoothed with a Gaussian with a FWHM of 4 pixels. The intensity scale is logarithmic. Sgr~A* and IRS13E are marked.
} \label{fig:ks_image_unnice}
\end{center}
\end{figure}

\subsection{SINFONI}\label{sinfoni}
Spectroscopy allows one to obtain a secure identification of spectral types and to measure radial velocities. 
We use spectra obtained with the integral field spectrometer SINFONI \citep{Eisenhauer_etal2003, Bonnet_03} at UT4 of the VLT\footnote{We use SINFONI data from ESO program 183.B-100.}. We analyze high-quality H+K-band laser and natural guide star data from May 21 and 23, 2009. The data from the latter date have a Strehl ratio of 28\% at $2.11\,\mu$m and a total of $2400\,s$ exposure. The spatial sampling is $12.5\,$mas $\times$ $25\,$mas pixel$^{-1}$, the spectral resolution is 1500. 
We apply the standard data reduction for SINFONI data, including detector calibrations (such as bad pixel correction, flat-fielding
 and distortion correction) and cube reconstruction. The wavelength scale is calibrated by means of emission line lamps and finetuned with atmospheric OH lines. The remaining uncertainty corresponds to typically (less than) $10\,$km/s. 
 
\section {Astrometry}\label{astrometry}
\subsection{Image processing}
One of the main aims of this study is to characterize the faint objects in the core of IRS13E, the velocities of which are unknown. We base our analysis on deconvolution of the images, which requires the determination of the PSF for each image.

For creating a PSF we use starfinder version 1.2 \citep{Diolaiti_etal2000_StarFinder}.
We extract the PSF from ten to 15 stars brighter than $m_{\mathrm{Ks}}=13.5$ and symmetrically distributed around IRS13E up to a distance of 4''. 
We then use the Lucy-Richardson algorithm \citep{Lucy_74}. Depending on the image quality, we choose between 6000 and 20000 iterations.  We are confident that the LR algorithm applied in this way also mostly correctly handles  extended sources.
For example, the extended object IRS2L \citep{Blum_96} contains only one maximum after deconvolution, and a large fraction of the flux remains detached from the central maximum. Relatively faint sources which are in the same distance from a bright star as the radius of the PSF are sometimes influenced by deconvolution rings. Most (and all important) sources are either clearly outside or clearly inside of these rings. 
We do not smooth the result of the deconvolution at all and find stellar positions by centroids around local maxima. 

We deconvolve all H- and Ks-band images. We then select by visual inspection the images for the subsequent analysis, on which at least some of the fainter objects are visible. These are 61 Ks-band images and 18 H-band images. This subjective quality selection is nearly identical to a Strehl ratio cut of 9\% in both bands. 

\subsection {Object detection} \label{subs:detec}
Some care is needed to distinguish between real and spurious sources since we are dealing with a crowded field imaged at the diffraction limit. Our procedure is as follows. We start by measuring the positions of the local maxima on the quality-selected, deconvolved images. We treat each frame individually, such that we can neglect maxima that are unlikely to correspond to a real source. These are for example maxima only 
marginally brighter than the noise floor or the background from the minispiral, or maxima that apparently are affected by deconvolution artifacts.

\begin{figure}
\begin{center}
\includegraphics[width=0.99\columnwidth]{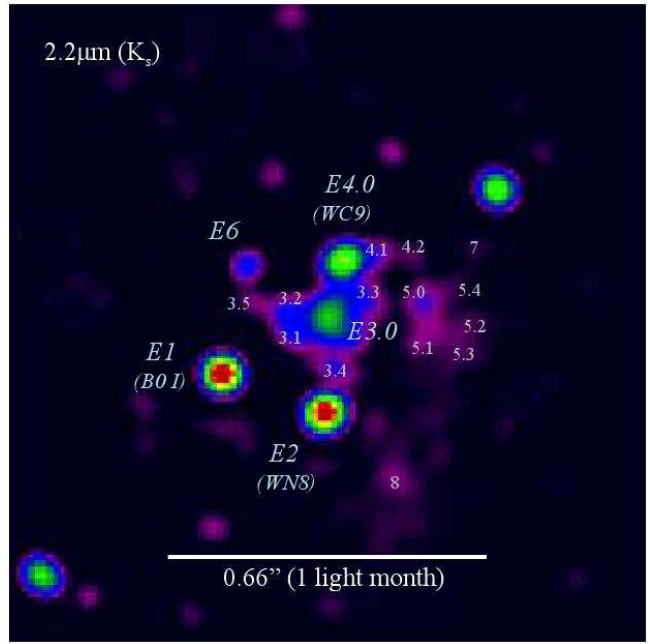}
\caption{Coadd of four Lucy-deconvolved Ks-images of IRS13E between 2002 and 2007, smoothed with a Gaussian with a FWHM of 2 pixels. The intensity scale is logarithmic. The sources which we consider real are marked with numbers. Nearly all of these objects are also found in H-band images.
} \label{fig:ks_image}
\end{center}
\end{figure}

We consider an object to be well-detected if it is found in either at least three images in both H- and Ks-band, or in at least ten Ks-band images. This is a robust criterion, since the objects are present in images that sample a wide range of PSF shapes. 
According to these criteria, we detect 19 sources in the core ($r\lesssim 0.3''$) of IRS13E (figure~\ref{fig:ks_image}).
The rms position deviation of these sources to a linear fit after transformation into
the astrometric reference frame to the Ks-band data is  $7\,$mas in median.  Thus all measured positions of the 19 sources lie within an area of $4000\,$mas$^2$, only 1/50 of the area of the cluster core. Therefore the sources are not random fluctuations.

We test by means of simulated images that indeed our detection procedure is robust. Using a variety of PSFs we find that it is only possible to continue to obtain the same sources after deconvolution if indeed these sources had been put into the simulation.

\subsection{Proper motions} \label {proper}
Given the pixel positions, we go to an astrometric coordinate system by constructing a full linear transformation using 
ten bright, isolated stars from the astrometric list of \citet{Trippe_08} close to IRS13E. Proper motions are in turn obtained by fitting the positions as a function of time by linear functions. Since we are using many position measurements for the fitting, we simply use an unweighted fit. The errors
for any given object are set to the scatter of the data around the linear motion model. For the fit, $5\sigma$-outliers are rejected. There are only two such cases in H and Ks for all the objects. The absolute position towards Sgr~A* has an additional uncertainty of about 2 mas \citep{Reid_07, Trippe_08, Gillessen_09} corresponding to a relative error of $\approx0.05$\% which safely can be neglected in our analysis. 

In the L'-band the exclusion of $5\sigma$ outliers becomes important, and the errors are mostly larger.
The L'-band velocities differ significantly from the velocities obtained from the other bands for the three brightest stars. The differences can be as large as $16\sigma$ and $2.5\,$mas/yr. 
The differences are probably caused by L'-band dust emission that is important around IRS13E. 
The L'-band velocities are less useful and we neglect them in our analysis.

The proper motion fits are presented in table~\ref{tab:one_motion} and figure~\ref{fig:positions1}. For the bright stars E1, E2 and E4.0 we also fit second order polynomials to the stellar positions as a function of time to test for accelerations. The accuracy reached  for the three brightest stars starts to constrain the mass of a putative IMBH. 

For most objects the velocities derived from H- and Ks-band data are
consistent. The biggest discrepancy happens for E4.0 that shows a velocity difference of 16 km/s or $3.7\,\sigma$ in R.A. The value for the R.A.-velocity in Ks lies between the velocities obtained from H- and L'-band data. Therefore, we assume that dust perturbs the position of this star also in Ks-band, not only in L'-band. 
Consequently we only use the astrometry from H-band data for this star, and we also don't trust the marginally significant acceleration in Dec. direction.

Similarly, for the faint or confused objects (E3, E4.1, E4.2, E5, E7, E8) we adopt the measurements derived from the band where the source is more prominent compared to the local background level. For the relatively blue objects E3.5, E4.0 and E4.2 this is the H-band, for the other ones Ks-band. For the unconfused objects E1, E2 and E6 we use the weighted average of the two bands.
 
\begin{deluxetable*}{llllllll} 
\tabletypesize{\scriptsize}
\tablecolumns{8}
\tablewidth{0pc}
\tablecaption{Proper motions and acceleration limits of the objects in IRS13E \label{tab:one_motion}}
\tablehead{Object & band  & R.A. & v$_{\mathrm{R.A.}}$& Dec. & v$_{\mathrm{Dec.}}$ &a$_{\mathrm{R.A.}}$ & a$_{\mathrm{Dec.}}$ \\
&&[as]& [km/s] & [as]&[km/s]&[$\mu \text{as/yr}^2$]& [$\mu \text{as/yr}^2$]}
\startdata
E1 & H and Ks &-2.958& -142.5 $\pm$ 1.3 & -1.645 & -105.5 $\pm$ 0.7 & 22 $\pm $ 20 & 11 $\pm $ 14\\
E2 & H and Ks & -3.171& -249.2 $\pm$ 1.1 & -1.732 & 23 $\pm$ 0.7 & 20 $\pm $ 19 & -14 $\pm $ 16\\
E3.0 & Ks & -3.184  & -82 $\pm$ 9 & -1.528 & 8 $\pm$ 9\\
E3.1 & Ks & -3.126 & 19 $\pm$ 39 & -1.562 & 19 $\pm$ 43\\
E3.2 & Ks & -3.093& -74 $\pm$ 38 & -1.525 & 84 $\pm$ 24\\
E3.3 & Ks & -3.252  & -262 $\pm$ 17 & -1.491   & 2 $\pm$ 15\\
E3.4 & Ks & -3.203 & -97 $\pm$ 20 & -1.632 & -23 $\pm$ 27\\
E3.5 & H & -3.026 & -12 $\pm$ 51 & -1.501 & -126 $\pm$ 48\\
E4.0 & H & -3.210  & -227 $\pm$ 4 & -1.409  & 26 $\pm$ 4 &  80 $\pm $ 70 & -30 $\pm $ 80\\
E4.1 & Ks & -3.281 & -211 $\pm$ 21 & -1.410 & 206 $\pm$ 18\\
E4.2 & H & -3.364  & -25 $\pm$ 43 & -1.402 & -80 $\pm$ 27\\
E5.0 & Ks & -3.382 & -136 $\pm$ 12 & -1.502 & 175 $\pm$ 14\\
E5.1 & Ks & -3.391 & -87 $\pm$ 38 & -1.569 & 42 $\pm$ 34\\
E5.2 & Ks & -3.474 & -340 $\pm$ 38 & -1.541 & 3 $\pm$ 28\\
E5.3 & Ks & -3.458  & -139 $\pm$ 39 & -1.607 & 163 $\pm$ 30\\
E5.4 & Ks & -3.446  & -68 $\pm$ 38 & -1.469 & 30 $\pm$ 34\\
E6 & H and Ks & -3.010  & -133 $\pm$ 7 & -1.423 & 23 $\pm$ 7\\
E7 & Ks & -3.495 & -227 $\pm$ 29 & -1.391 & 167 $\pm$ 28\\
E8 & Ks & -3.320  & -118 $\pm$ 18 & -1.866 & 95 $\pm$ 20\\
\enddata
\tablecomments{The positions refer to reference epoch 12-05-2005 and are measured relative to Sgr~A*. The dominating uncertainty of the position is caused by the coordinate uncertainty of about 2 mas \citep{Reid_07, Trippe_08, Gillessen_09}. }
\end{deluxetable*}

\section {Characterization of the objects}\label{photometry}

In order to address the question of whether IRS13E hosts an IMBH, it is important to know the nature of its constituent objects. Only stars can be used as fiducial tracers for a gravitational potential, and an overdensity of objects compared to the stellar background is only meaningful if these are actually stars. In this section we first describe the combined photometric and spectroscopic method to construct a SED. Then we use it to constrain the nature of the individual objects. We present the results for the most important objects here and leave for the appendix a description of the details of the method and the results for the remaining objects.

\subsection {Method}

For each object we construct a spectral energy distribution (SED), ideally ranging from H- to L'-band. For this we use
broad- and narrow-band images plus the two spectro-imaging data cubes in the combined H+K-band.

Where possible, we identify the objects in the SINFONI data by their spectral features as late- or early-type stars. In this way we show that E1, E2, E4.0 and E6.0 are stars. For the other objects we construct SEDs from a deconvolved SINFONI cube and the narrow-band imaging data (appendix~\ref{sec:ob_seds}). We then estimate whether a stellar component is required to explain the SED. 

Following the approach of \citet{Maillard_04}, we consider models composed of an extinction parameter and either one or two temperatures corresponding to a star and dust. We derive the temperature of the stellar component from the lines in the spectrum when possible. Otherwise we assume a temperature of $9480\,$K (corresponding to the spectral type A0V) for the star. This is sufficient, since the shape of the SEDs  in our wavelength range depends only weakly on the assumed temperature for the stellar component of the SED model.

We use the extinction law of \citet{Draine_89} shortward of Brackett-$\beta$ and
$A_{Ks}/A_{L'}=1.75\,$ \citep{Lutz_96}. When fitting models to the less well measured SEDs, we only use extinction values that differ less than $\Delta A_{Ks}=0.3$ from the value for E1. This range seems adequate given the extinction map of \citet{Schoedel_09b} and our result does not depend strongly on this assumption.
We neglect the nebular lines originating from the interstellar gas. At our precision their contributions to the flux are negligible

For the bright early-type stars we have well-measured spectra such that an explicit comparison with atmosphere models becomes viable.
We use the code CMFGEN \citep{Hillier_98} to derive the main parameters of the stellar atmospheres of E1, E2 and E4.0. The code computes non-LTE atmosphere models including wind and line-blanketing. We proceed as in \citet{Martins_07}.  A slight improvement in the method is the use of iterations on the hydrodynamic structure to ensure a better consistency between the atmosphere model per se and the density/velocity structure \citep{Martins_09}. The derived stellar parameters rely on line ratios and photometry. In addition, we derive the 
extinction by fitting the shape of the infrared SEDs. Hence, for each star, 
we have computed several models with different luminosities and looked 
for the best combination of luminosity, extinction and dust emission.

\subsection {Individual Objects}

This section describes the nature of the most important objects in IRS13E. For a description of the other objects, see appendix~\ref{sec:objects}. 
Table~\ref{tab:spec_properties} presents the values of the model parameters for all objects, for which our data allow an identification.

\begin{figure}
\begin{center}
\includegraphics[width=0.707\columnwidth,angle=-90]{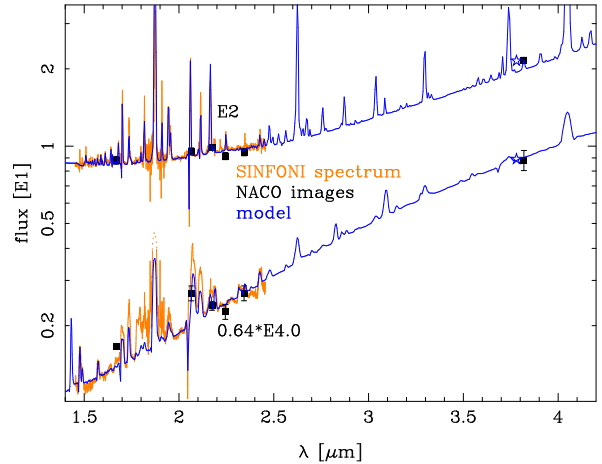}
\caption{H+K-band spectrum and photometry of the two WR-stars in IRS13E: E2 (WN8) and E4.0 (WC9). We fit full atmosphere models plus blackbodies with dust-like temperatures to the observed spectra and L'-band flux points. The asterisks denote the average flux of the models in L'-band, to which the measured flux needs to be compared.
} \label{fig:spec_E2}
\end{center}
\end{figure}

\subsubsection{IRS13E3.0}
\label{sece3}

We resolve E3 on the deconvolved images into six stable sources and name them E3.0, E3.1, E3.2, E3.3, E3.4 and E3.5. This section describes the brightest component E3.0, which likely is extended or multiple at our resolution.
Its flux increases significantly by 0.25 mag between 2002 and 2008 in H- and Ks-bands, the main increase happens between 2004 and 2006. 
This is not caused by confusion. 
 
\begin{figure}
\begin{center}
 \includegraphics[width=0.79\columnwidth,angle=-90]{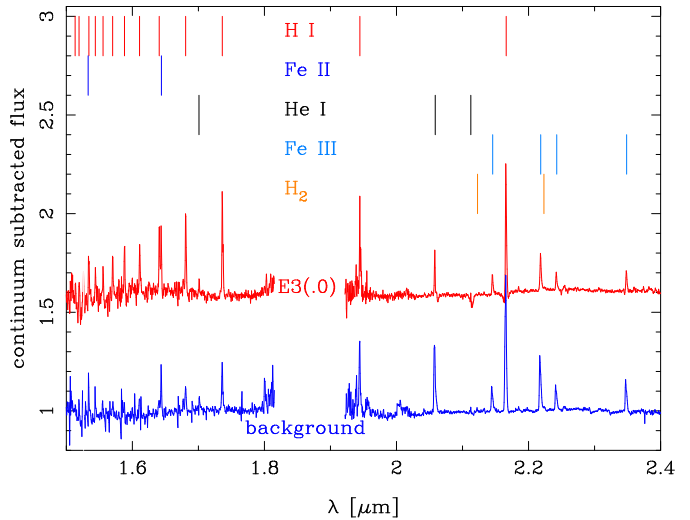}
 \caption{H+K-band spectrum (red) of the brightest component of IRS13E3: E3.0. For comparison the spectrum of the nebular background is shown, too (blue, shifted). In the upper part, the line positions for various transitions are marked.
The weak CO band heads in the background spectrum are caused by stray light of late-type stars.
The weak absorption feature of E3.0 around Brackett $\gamma$ and Helium are likely caused by the slightly imperfect subtraction of the stellar emission lines.
} \label{fig:lines_E3.0}
\end{center}
\end{figure}

We extract a spectrum from the SINFONI data at the position of E3.0 (figure~\ref{fig:lines_E3.0}). This spectrum also contains flux from the other five E3 sources and the extended background, but E3.0 completely dominates due to its brightness. 

No stellar line (absorption line, or broader emission line) is visible in E3.0, but many narrow emission lines from H, Fe and He are present. Nearly all these lines are very similar to the surrounding nebular background. A few lines like H$_2$ at $2.122\,\mu$m and $2.223\,\mu$m are only present in the background. Most identifiable lines (HI, HeI, FeII, FeIII) are known from the minispiral \citep{Lutz_93} or are other hydrogen lines. 

Stellar absorption lines of fainter stars are likely too weak for being detectable against the strong narrow emission lines of the surrounding gas. However, that there are any WR-stars in E3.0 seems unlikely, since the strong, broad emission lines of such stars would be visible. For example, E4.0 (a WC9 star) scaled down to $m_\mathrm{Ks}=15$ would be detectable against the narrow emission lines. Given that a WR-star cannot be fainter than $m_\mathrm{Ks}=16.5$ \citep{Crowther_07}, only a few types of WR-stars could be hidden in E3.0. Also, E3.0 does not resemble a dusty WR-star (of which E4.0 is one), in which case the lines would be diluted but still visible.

It is worth noting that also the background emission is well determined from our data. 
The seeing halo of E3.0 is of no concern since the lines in the background are stronger in relative terms. 
The extinction appears to be similar in the whole region, since we do not measure any change in the relative line strengths for lines from one ionization state of an element over the spectral cube. 
In contrast, the line strength ratios of different elements differ between the background and E3.0 (see for example the HI, FeII pseudo doublet at $1.64\,\mu$m in Figure~\ref{fig:lines_E3.0}). These spatial changes in the spectral properties appear to be gradual, consistent with a gaseous nature of E3.0. They are strongest at the position of E3.0, but they do not correlate in general with the positions of the red sources in IRS13E.

The radial velocity of E3.0 determined from the Brackett-$\gamma$ line is $v_\mathrm{LSR}\approx-25\,$km/s, measured at the position of the continuum peak of E3.0.
 \citet{Paumard_06} measured a velocity of $v_\mathrm{LSR}=87\pm 20\,$km/s for E3 but did not identify the type of star. Probably \citet{Paumard_06} measured an average of E2 and E4.0 instead of the E3 velocity. 
 
\begin{figure}
\begin{center}
\includegraphics[width=1.34\columnwidth,angle=-90]{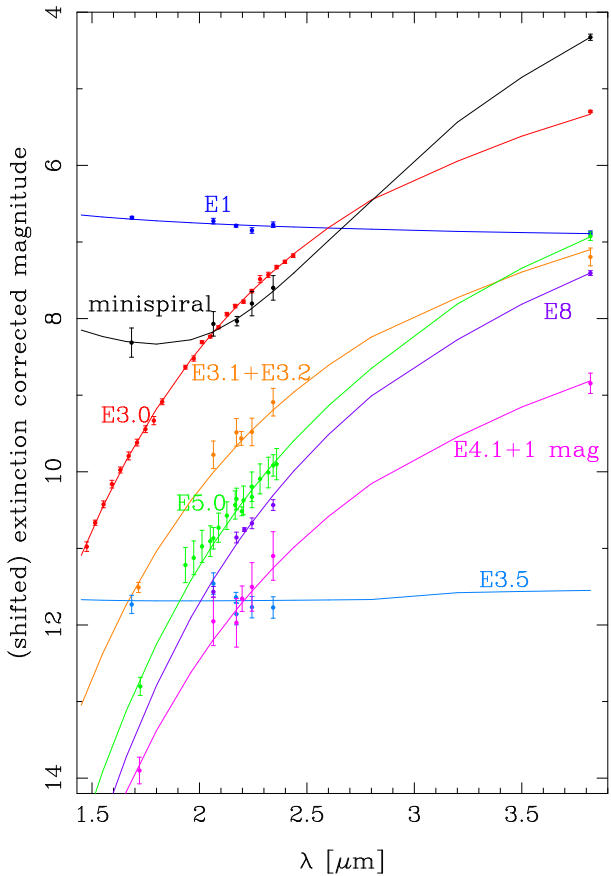}
\caption{SED data and model fits for E1, E3.0, E3.5, E4.1, E5.0, E8, the minispiral and the sum of E3.1 and E3.2. All data and fits are offset from each other for clarity. For all sources the extinction correction of E1 is applied. E1 and E3.5 are fit well by single blackbodies of stellar temperature. 
The objects E3.0, E3.2+E3.3, E4.1, E5.0 and E8 can be fit as single blackbodies with a dust-like temperature. Stellar components do not improve these fits.
The SED of the minispiral shows two blackbodies: A warm dust component and stellar stray light. 
} \label{fig:many_SED}
\end{center}
\end{figure}

Given the photometric variability of E3.0, we fit the SED based on the narrow-band data set from 2004 and the one based on the SINFONI spectroscopy from 2009 separately. Using the latter we obtain $A_\mathrm{Ks}=3.89 \pm 0.16$ and a warm dust component of $T=970 \pm 17\,$ K, see table~\ref{tab:spec_properties} and figure~\ref{fig:many_SED}. The $1\sigma$ flux limit on the stellar component is  $m_\mathrm{Ks}=16$. Using the earlier imaging data yields consistent results. Also in this case, no stellar component is needed, however the limit is less constraining due to the smaller wavelength coverage. The extinction value $A_\mathrm{Ks}=3.5 \pm 0.7$ is consistent with the former one. 
We also obtain an independent extinction estimate of $A_\mathrm{Ks}=3.44\pm0.15$ from the Brackett series of E3.0 by using the relative line strengths of \citet{Hummer_87} and the electron densities and temperatures from \citet{Lutz_93}. 
Our different extinction estimates appear to be sufficiently consistent with each other. Using a value around $A_\mathrm{Ks}=3.4$ in conjunction with the 2009 data yields that the magnitude of a possible stellar component is less than $m_\mathrm{Ks}=18$.

We also compare the spectrum of E3.0 with those of young stellar objects (YSO). Such objects are not expected to reside in the central parsec, because the gas density there is much too low for star formation \citep{Christopher:2005p1402,Bonnell_08}. \citet{Eckart:2004p1170} and \citet{Muzic_08} speculate that some of the objects of IRS13N nevertheless could be YSO, arguing that shocks could allow  star formation to take place. 
The combination of color and luminosity for E3.0 argues against a typical Herbig-Be-star. Also, no distinctive feature of a massive YSO is visible in E3.0. Except for the Brackett lines, the lines observed in E3.0 are unknown (e.g. FeIII)  or weak (e.g. He) in massive YSO. A massive YSO does not necessarily show other lines apart from Brackett-$\gamma$ in the K-band, but often CO-band heads, H$_2$, or the Pfund series  in emission can be visible \citep{Bik_06,Hernandez_08}. Such a clear indication for a YSO is not observed in E3.0, nor in other parts of IRS13E. Furthermore, there is no additional local extinction towards E3.0 while YSOs show locally enhanced extinction. 
We conclude that it is unlikely that E3.0 is a YSO. 

The source E3.0 in the H-band has a velocity similar to the the radio source in IRS13E \citep{Zhao_09}.
Therefore the H-band emission is likely due to dust associated with the gas, which causes the radio continuum emission.
Hence, also qualitatively a higher extinction value (like for example expected in YSO) seems not to be needed.  

\citet {Maillard_04} found two similarly bright components E3A and E3B, which is different from our result at higher resolution.
Their additional stellar components do not fit our SED for E3.0. 
Our flux at $1.48\,\mu$m is a factor $\approx1.8$ lower than the sum of the flux models for E3A and E3B, but at $2.18\,\mu$m our flux is only a factor $\approx1.15$ lower.
The hint for the stellar component in  \citet {Maillard_04} is only based on the $1.6\,\mu$m-flux measured in the filter F160W from NICMOS.
 Since this is a very broad filter, and E3.0 is a very red source, the central wavelength of the filter is considerably bluer than the effective wavelength. Hence, using the central wavelength will overestimate the flux. We speculate that this effect made the stellar component necessary. Also, the amount of stellar flux contributing at that wavelength to the total flux is small, and hence a small measurement error might change the conclusion from \citet {Maillard_04}.

To summarize; the potential stellar component in E3.0 is fainter than most known WR-stars. A main sequence star in E3.0 would need to be fainter than B1.5V. This seems unlikely, however, since these stars do not produce dust. 
With three different methods we obtained consistent extinction values, similar to the extinction towards E1 and E2.  We conclude that E3.0 is neither a star nor a YSO, but rather a concentration of warm dust and gas. 

The source E3.0 is not the only dust source of its brightness in the GC. For example, the source IRS2L resembles E3.0 in many respects and also shows extended H-band emission.

\subsubsection{E7, E8 and the surrounding gas}
\label{e7e8gas}
The background around IRS13E is not only visible in gaseous emission lines (figure~\ref{fig:lines_E3.0}) but also in the continuum. 
The deconvolution artifacts hamper any photometric measurement of this continuum. Instead, we use the objects E7 and E8 (figure~\ref{fig:ks_image}) that are less affected. These objects are farther away from the center of IRS13E than the other sources and likely do not belong to IRS13E but to the background. Spectroscopically, E8 appears to be indeed a local overdensity in the gas.
 
E7 and E8 are not detected in the H-band. We fit a blackbody model to each of the two objects, fixing the extinction to its upper bound at $A_\mathrm{Ks}=3.9$. We find that no star is needed to explain the SED of these two objects. We obtain a temperature of $T=692 \pm 55\,$K for E7 and of $T=732 \pm10\,$K for E8. Thus, E7 and E8 are colder than most objects in IRS13E. 

We also detect three faint stars ($m_\mathrm{Ks}\approx15$) in the SINFONI data at the approximately same distance ($\approx 450\,$mas) to  E3.0. All these stars show CO band heads. \citet{Buchholz_09} correctly identify the star with the strongest CO-band heads as late-type, the others are misidentified as early-type stars. 
The radial velocities of these stars are different from the radial velocity of IRS13E. Thus, the stars differ from IRS13E not only in proper motion \citep{Trippe_08} but also in stellar type and radial velocity. We conclude that they are fore- or background objects.

The background around IRS13E is part of the minispiral \citep{Lo_83,Paumard_04,Zhao_09}.
The minispiral is visible in our L'- and Ks-band data.  In order to measure its SED, we select seven boxes of $\approx0.2$'' diameter far away from bright stars at a distance of around 1.5'' to IRS13E. We exclude the narrow-band data around Brackett-$\gamma$ because of the strong emission line there.  
We calibrate the flux from the narrow-band filters  by fitting these flux values as a linear function of wavelength and setting equal the flux at $2.18\,\mu$m to the measured Ks-band flux. We fit two blackbodies to the SED data (figure~\ref{fig:many_SED}). One accounts for the stellar stray light, the other represents the dust. The fits for all fields are similar.
The average extinction is $A_\mathrm{Ks}=3.64\pm0.55\pm0.56$, consistent with the extinction value derived for E1. 
The average dust temperature is $T=585\pm 22 \pm 33\,$K. Thus, the minispiral is colder than the dust in IRS13E. 

One of the seven fields contains the central sources of IRS13N \citep{Eckart_04,Moultaka:2005p2410,Muzic_08}. In this case the fit yields $A_{\text{Ks}}= 3.7 \pm1.6$ and $T= 664 \pm 174\,$K. This argues against a highly extincted object (like a YSO) as these authors had proposed. The object IRS13N$\beta$ is within the field of view of our SINFONI data. We do not find any significant differences between its spectrum and the spectrum of the gas around IRS13E. 
Also, the velocity of IRS13N in L'-band \citep{Muzic_08} and at radio wavelengths \citep{Zhao_09} make it more likely that IRS13N actually consists of dust blobs. They appear denser than in other regions in the GC, but still they are much less prominent than E3.0. 
The common proper motion of these sources \citep{Muzic_08} could
also be explained by a recent formation of the dust clouds. 

\begin{deluxetable*}{lllccccc} 
\tabletypesize{\scriptsize}
\tablecolumns{8}
\tablewidth{0pc}
\tablecaption{Spectral properties of the objects in IRS13E \label{tab:spec_prop}}
\tablehead{ Name & data & spectral type & $A_\mathrm{Ks}$ & $T_1$ [K] & $T_2$ [K]&  star $m_\mathrm{1,\,Ks}$ & dust $m_\mathrm{2,\,Ks}$ }
\startdata
E1& NB, H, K, L & OB I & 3.64 $\pm$ 0.07 & 26000 & & 10.38 $\pm$ 0.06 &  \\
E2 & SINF, NB, H, K, L &WN 8 & 3.51  & 30000 & 712 & 10.43 & 14.74  \\ 
E3.0 & SINF, NB, H, K, L & gas & 3.89 $\pm$ 0.16 & 9480 & 970 $\pm$ 17 & $>16$ & 11.42 $\pm$ 0.05 \\
E3.1+E3.2& NB, H, K, L & gas & 3.9  & 9480 & 955 $\pm$ 24 & $>18.03$ & 13.24 $\pm$ 0.18 \\
E3.3& NB, H, K, L & gas & 3.9  & 9480 & 987 $\pm$ 54 & $>15.16$ & 13.65 $\pm$ 0.35 \\
E3.4& NB, H, K, L & gas & 3.9  & 9480 & 954 $\pm$ 49 & $>16.53$ & 14.15 $\pm$ 0.36 \\
E3.5& NB, H, K &  & 3.49 $\pm$ 0.36 & 4000-19000 & & 15.30 $\pm$ 0.26 & \\
E4.0& SINF, NB, H, K, L & WC 9 & 3.64 & 41000 &  1400 & 11.95 & 12.51 \\ 
E4.1& NB, H, K, L & gas & 3.9  & 9480 & 844 $\pm$ 24 & $>18.36$ & 14.38 $\pm$ 0.21\\
E4.2& NB, K, H &  & 3.9  & 4000-19000 & 517 $\pm$ 252 & 15.77 $\pm$ 0.11 & 16.64 $\pm$ 2.5 \\
E4.3&K &  &  &  &  &  & 15.48 $\pm$ 0.10\\
E5.0& SINF, NB, H, K, L & gas  & 3.4 & 9480 & 700 $\pm$ 8  & $>18.46$ & 14.08 $\pm$ 0.10\\
E5.1& NB, H, K, L & gas & 3.9  & 9480 & 947 $\pm$ 25 & $>17.75$ & 14.25 $\pm$ 0.14 \\
E5.2& NB, K, L & gas & 3.9  & 9480 & 642 $\pm$ 70 & $>15.6$ & 15.49 $\pm$ 0.10\\
E5.3& NB, H, K & gas & 3.9  & 9480 & 978 $\pm$ 524 & $>16.04$ & 15.1 $\pm$ 1.6 \\
E5.4& NB, K & gas & 3.4  & & 817 $\pm$ 474 & & 15.54 $\pm$ 2\\
E6 & NB, H & K3III & 3.68 $\pm$ 0.09 & 4300 & & 13.82 $\pm$ 0.10 & \\
E7 & NB, K, L&  & 3.9  & 9480 & 692 $\pm$ 55 &$>16.71$ &  16.02 $\pm$ 0.40\\
E8 & NB, K, L & gas & 3.9  & 9480 &  732 $\pm$ 10 & $>17.14$ & 14.51 $\pm$ 0.11 \\
\enddata
\tablecomments{We fit one or two blackbodies with temperatures $T_1$ and $T_2$ and one extinction parameter to our SED data.  We allow extinction values between $A_\mathrm{Ks}=3.4$ and $A_\mathrm{Ks}=3.9$ and clip it if needed at the respective edge of the interval. We derive the temperature of the stellar component from an atmosphere model or from the observed lines when possible. Otherwise the temperature is set to $T=9480\,$K.  
All limits are $1\sigma$-limits. In column two, SINF means the SED fitting results are based on SINFONI data, NB that the narrow band images are used.
}\label{tab:spec_properties}
\end{deluxetable*}

\section {Probability for an IMBH in IRS13E}
\label{disolving}

We now discuss the nature of IRS13E considering our NIR observations. There are four main explanations:
\begin{itemize}
\item IRS13E is a cluster in the process of dissolution, one of various possibilities suggested by \citet{Schoedel_05}. This is very unlikely, because it would get diluted extremely quickly. In the case of IRS13N with a similar dispersion as IRS13E \citet{Muzic_08} show  in their figure 2 that such an association is diluted quickly in the GC.  We estimate that after just 200 years one would not recognize anymore the overdensity of three young, massive stars of IRS13E as such.  We exclude this option.
\item IRS13E is a cluster bound by stellar mass. From our detection limits, we estimate that the total stellar mass of the cluster could at most be $2000\,M_\odot$. 
\item IRS13E is a cluster bound by additional mass, e.g. an IMBH.
\item IRS13E is a chance association. 
\end{itemize}

\subsection{Stellar surface density in IRS13E} 
A physical cluster leads to an overdensity in projection compared to the local background density. 
Detecting a local, significant overdensity in the stellar surface density would hence be a strong indication
that IRS13E is a cluster. From our data we measure the projected space densities of stars in IRS13E. 
\citet{Paumard_06} found 13 objects with $m_{\text{H}}<19.4$ inside a radius of 0.3'' in IRS13E.
Similarly, we find 15 objects in H- and Ks-band with $m_{\text{H}}<18.5$ inside a radius of $0.27''$.

\begin{figure}
\begin{center}
\includegraphics[width=0.707 \columnwidth,angle=-90]{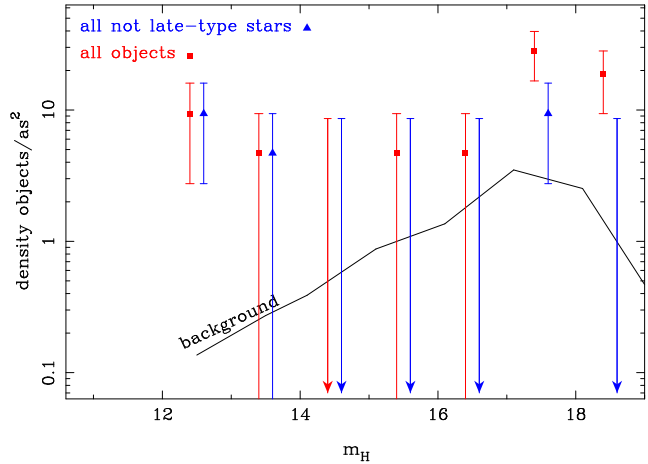}
\caption{Surface densities of objects in the core ($r=0.27$'') of IRS13E. The errors are Poisson errors. The upper limits are $1\sigma$-limits. Red: All objects. Blue: Cluster member candidates only. The red and blue data are offset slightly in the horizontal direction for clarity. The solid line shows the estimated background density for our data.
} 
\label{fig:_star_density}
\end{center}
\end{figure}

\begin{figure}
\begin{center}
\includegraphics[width=0.88 \columnwidth,angle=-90]{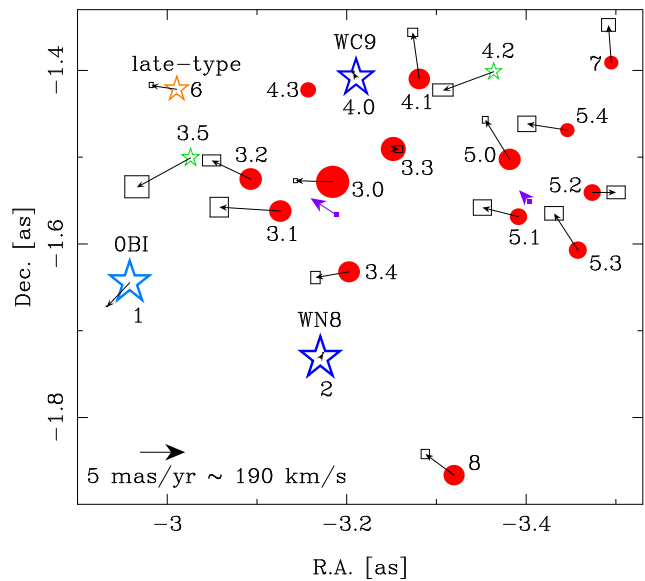}
\caption{Velocities of the objects in IRS13E relative to the mean motion of the two WR-stars E2 and E4.0. 
The near-infrared motions are marked with black arrows, the boxes denote the  1-$\sigma$ errors.
Stars are marked with asterisks (blue: young stars; green: stars without spectral identification). Dust clumps are marked by red disks. The sizes of symbols indicate the Ks-band flux.
The $13\,$mm radio motions (violet) of \citet{Zhao_09} are shown, too. 
} 
\label{fig:_v.objects}
\end{center}
\end{figure}

We determine the density of background stars in a non-confused region with a size of $5.15\,$sqas north of IRS13E for stars fainter than $m_\mathrm{H}=14$. Because  there the extinction is higher than towards IRS13E according to the extinction map of \citet{Schoedel_09}, we shift the magnitudes of all the stars in this field by 0.4 magnitudes to brighter magnitudes. For stars brighter than $m_\mathrm{H}=14$ we use the results from \citet{Genzel_03} to estimate the background density. We find that the density of objects in IRS13E is higher than in the surrounding field at nearly all magnitudes (figure~\ref{fig:_star_density}), if all objects are considered. However, only for six of these objects a star is required to explain the observed SED. 
For the other objects, a possible stellar component needs to be fainter than about $m_\mathrm{Ks}\approx15.5$.
This corresponds to a stellar surface density of $26\,$as$^{-2}$, which is five times higher than the background.
However, E6 (figure~\ref{fig:_v.objects}) is not a member of the cluster, given its age well above $6\,$Myr. 
Furthermore, the two faint stars E3.5 and E4.2 have magnitudes consistent with the red clump and hence are likely also older than the bright stars. In the magnitude bin of E3.5 and E4.2, the background population accounts already for one of the two stars. Thus, the overdensity in this bin is insignificant, too.

We conclude that the magnitude integrated overdensity of objects in IRS13E (with a chance probability  of $2 \times 10^{-5}$, \citet{Paumard_06}) is caused by dust clumps and not by stars. This overdensity cannot be used as evidence for a cluster. Nevertheless, the concentration of the three bright early-type stars in IRS13E remains puzzling, an argument which we analyze quantitatively in section~\ref{3_stars}. 

\subsection{The structure of IRS13E}\label{sec:struc}
A significant number of dust clumps in the GC are found far away from stars \citep{Muzic_07}. But most of these dust clump are invisible in Ks-band and none of them is visible in H-band. In the H-band the dust blobs in IRS13E are among the brightest non-stellar structures in the central parsec. 
The dust clumps close to E3.0 are nearly twice as hot as the minispiral and also denser ($\approx 4 \times$ for E3.0, estimated  from surface brightness and temperature) than the minispiral. The brightest and hottest dust blobs (around E3.0) are located between the two bright WR-stars E2 and E4.0. The star E2 is the brightest of all WR stars in the GC and E4.0 is a dusty WR with the hottest dust in IRS13E.
  
Therefore E4.0 might be the most important dust source in IRS13E. But it seems unlikely that E4.0 is responsible alone for the bright dust blob E3.0, 
given that it does not move away from E4.0, while the motions of the blobs E4.1 and E4.3 are consistent with being outflows from E4.0.
Instead, we assume that the winds of E2 and E4.0 collide in the E3 region and cause the bright dust blobs in this way. For this model it is not necessary that the dust originates from the stars but only that it is heated by the two stars. The fact that the velocities of these blobs are not pointing towards either of the two WR stars supports this picture. Also, the eastward component of the R.A. motion of the blobs increases from west to east as expected in this model.
Accordingly, E2 and E4.0 would not only be close in projection, but would indeed have a small 3D distance.

A potential weakness of that model is that the motions of the blobs E3.3 and E4.1 do not seem to fit. But this is only a worry if no other forces than repulsion from the stars act on the winds. For example, the presence of the gas from the minispiral might decelerate the winds. A second problem might be that it is not clear whether the conditions for dust formation are satisfied, given that our system is very different from WR140 \citep{Monnier_02}. 

Most likely E1 does not contribute to the dust production. It is further away and supergiants have weaker winds than WR-stars. Furthermore, the velocities of some dust clumps are pointing towards E1. Therefore E1 can, but needs not to be close in real space to the WR-stars.

The dust clumps in E5, E7 and E8 fit less well into our simple model of two WR stars because of their velocities. They are also mostly colder than the dust clumps in E3 and E4, but still warmer than the minispiral. In addition, they are farther away from E2 and E4.0. Likely the minispiral is more important for their formation.

\subsection {The overdensity of young, massive, early-type stars}\label{3_stars}

Because most of the fainter objects in IRS13E are dust clumps, the most prominent feature of the potential cluster IRS13E is the concentration of the three bright early-type stars. 
Therefore we quantitatively evaluate the argument of \citet{Maillard_04}, that this association cannot be a chance occurrence.
On the one side, the number of stars is very small, and one could argue that IRS13E cannot be a significant overdensity. On the other hand, five or even six phase space coordinates of these stars are similar.

We calculate the probability that a star association like IRS13E occurs randomly, i.e. a pseudocluster, by means of Monte Carlo simulations. IRS13E cannot be part of the clockwise disk of young stars \citep{Paumard_06}.
Consequently, the size of the sample of stars considered is chosen to equal the observed population of those O- and WR-stars that are not part of the clockwise disk, which yields 96 stars with $r<10''$. We assume that as many of the counterclockwise moving stars are in a disk-like configuration as suggested by \citet{Bartko_09}; the rest is on randomly oriented orbits. The disk thickness and eccentricity distribution  of the counterclockwise disk are not well known. The thickness is likely larger than 10$^{\circ}$ (the value for the clockwise disk) and smaller than 25$^{\circ}$, because a thicker disk would be invisible \citep{Bartko_09}. The eccentricity distribution is more uncertain. For each realization we have a set of eccentricities with an rms width of 0.22  \citep{Bartko_09} and a median eccentricity which is varied for each realization. The median values are chosen such that they are distributed symmetrically around the mean eccentricity of the clockwise disk, in the range from 0.17 to 0.60 \citep{Paumard_06, Bartko_09}. We use these eccentricity distributions for the counterclockwise disk and for the random stars. The impact of the width of the eccentricity distribution on the derived probabilities is small compared to influence of the median eccentricity.  
The distribution of semi-major axes is chosen such that the final surface density profile resembles the one observed for the stars that are not part of the clockwise disk. A change of the semi-major axis distribution compatible with the errors of the observations has nearly no influence on the derived probabilities.
The actual number of stars is varied, too. On purpose, we don't restrict the sample to the stars at least as bright as the IRS13E members. Fainter O-stars are barely less massive, and dynamically they would not behave any differently \citep{Zwart_07}. 

The criteria for finding a cluster from the simulated stars are motivated by the observed properties of IRS13E. 
If anywhere in the simulated set three stars are found for which the following four conditions are met, the set is marked as containing a pseudo-cluster of three stars. The following cut values are the measured values plus the $1\sigma$ errors: We require that $r_\mathrm{cluster,\, 2D}\leq 0.19''$ and
that the proper motion dispersion $\sigma_\mathrm{v,\,2D} \leq 94\,$km/s. Furthermore, we demand that for any two stars of the three $r_\mathrm{3D}\leq0.46''$ and for any (other) two stars of them $\Delta v_z \leq76\,$ km/s,  A pseudo-cluster of two stars is defined by the criteria $d_\mathrm{2D} \leq 0.33''$,  $\sigma_\mathrm{v,\,2D}\leq 23\,$km/s,  and $d_\mathrm{3D}\leq0.46''$. In the following we give the probabilities to find such configurations, and in brackets we give the values if the requirement that two stars need to be close in the $z$-coordinate is omitted.

\begin{figure}
\begin{center}
\includegraphics[width=0.65\columnwidth,angle=-90]{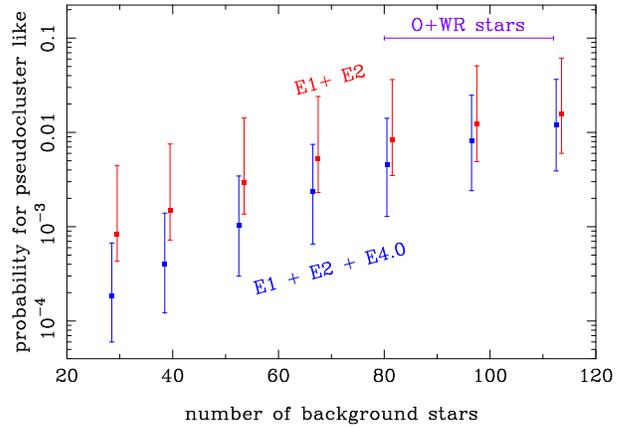}
\caption{Probability for observing a cluster of two (red) or three (blur) early-type stars that resemble IRS13E in a background population of given sample size. The errors are derived from assuming different distributions of orbital elements for the simulated stars.  The red and blue data are shifted for clarity horizontally. The background population consists of the O/WR early-type stars outside of the clockwise disk. The violet bar marks the range applicable for the GC.
} \label{fig:_bright_probability}
\end{center}
\end{figure}

By randomly drawing a stellar distribution many times and varying the assumptions on the dynamical properties of the O/WR stars population, we find that a triple of stars matching the criteria for a pseudo-cluster of three stars out of 96 has a chance probability of  0.2 \% to 2\% (0.6\% to 4\%) with the most likely probability being 0.8\%. Finding a pair of stars out of 96 that matches the criteria for a pseudo-cluster has a chance probability of  0.4\% to 5\% (4\% to 14\%, figure~\ref{fig:_bright_probability}). 
Demanding in addition that E2 and E4.0 have similar masses lowers the chance probability for such a pair to 0.1\%, since only 28 of the 96 stars from which we draw the pseudo-clusters have masses comparable to that of E2 and E4.0.

The dominant error source here are the assumptions on the dynamical properties of the O/WR stars population, as indicated by the ranges of values in the previous paragraph. The uncertainties of the probabilities induced by the measurement errors of the proper motions are very small. They lead
to relative errors of the probabilities well below 10\%. Also the influence of
 $\delta v_\mathrm{z}$ on the probability is small with 13\% (relative error). 
 We conclude that the occurrence of a pseudo-cluster similar to IRS13E is not expected, but still not 
impossible\footnote{In the clockwise disk one also finds a comoving group of stars: The IRS16-SW group \citep{Lu_05}. This group is less concentrated than IRS13E.}.

We also note that E2 and E4.0 could be a binary in an orbit with a pericenter distance of more than $1.3\,$pc (see section~\ref{e2.e4}). Heavy binaries indeed often consist of two similarly massive stars \citep{Kobulnicky_07}. On the other hand, the hypothesis that all three stars are physically connected and reside outside the central parsec seems improbable since such young, massive stars actually are very rare in the Galaxy. And regardless of the location of IRS13E along the line-of-sight, the high velocity dispersion would require additional, unseen mass to bind the cluster. 

\subsection{Cluster mass from velocity dispersion}
\label{clustermass}
Because the probability for IRS13E being a chance association is relatively small, it is reasonable to exploit the idea that IRS13E indeed is a cluster. Therefore we estimate the mass needed to bind the cluster by analyzing the velocity dispersion.

\subsubsection{Assuming E1, E2 \& E4.0 are cluster members}
\label{three}
Classical mass estimators are very uncertain when applied to a sample of three stars only. Hence, we simulate clusters of three stars around a central mass in a Monte Carlo fashion for deriving the probability distribution for the value of the mass. 
For this one needs to marginalize over many parameters, mainly the 18 orbital elements of the three stars.
Since we assume here that IRS13E is a cluster, we normalize the probability such that the total probability of clusters accepted over the full mass range is 1. 

We assume that all angles and phases are randomly distributed and that the eccentricities follow a thermal distribution $n(e)\propto e$.  
The semi-major axes $a$ are drawn from $a=0.4'' \times p^{0.9}$, where $p$ is a random number in [0,1]. This makes the distribution of projected distances similar to the observed one. 
In this way we simulate $5\times10^8$ clusters around different IMBHs with masses between $300\,M_{\odot}$  and $3\times 10^5\,M_{\odot}$ in steps of $\Delta M/M=0.15$. 

The peculiar property of IRS13E is that one star has a significantly different velocity compared to the other two. Thus, we choose from the simulated clusters the ones, which have nearly the same velocity differences as the observed cluster: Between any two stars this is $| V_\mathrm{2D,\,sim}-V_\mathrm{2D,\,obs}|<10\,$km/s and between two of them additionally $|V_\mathrm{z,\,sim}-V_\mathrm{z,\,obs}|<39\,$km/s.
We neglect the exact spatial distribution of the stars in IRS13E. But on average the simulated stars from the chosen clusters are at similar positions as the observed stars. 

The uncertainties of the derived probability density due to the measurement errors and due to the assumptions on the semi-major axis distribution are small. Even though the exact shape of the distribution changes slightly, it will continue to be a concave function normalized to 1, with a broad maximum around $2\times 10^4\,M_\odot$.

We derive that the most likely mass is $M_\mathrm{IMBH}=16900\,M_{\odot}$ with a 68.3 \% confidence region from $4000$ to $59400\,M_{\odot}$. The probability distribution for $M_\mathrm{IMBH}$ is highly non-Gaussian, see black curve in Figure~\ref{fig:_tot_prob}.
The large uncertainty on the mass is due to the low number of stars. Nevertheless, the mass needed to bind IRS13E exceeds the mass of the three stars. 
Hence, if IRS13E is a cluster of three stars, it needs to be bound by additional mass that is not detected by our observations. This mass could be present in the form of an IMBH.

\subsubsection{Assuming only E2 \& E4.0 are cluster members}
\label{e2.e4}
If one assumes that E2 and E4.0 form an equal-mass binary, the measured velocity difference yields a mass for each star of $180\,M_{\odot}$. The two stars have almost
certainly a much lower mass, but the argument illustrates that an
IMBH with a mass similar to the equal-mass binary is sufficient to bind the two stars, too.
Since this mass is less than the limit on the stellar mass in IRS13E ($2000\,M_\odot$), one would not need an IMBH to explain IRS13E. 

The simple binary hypothesis is problematic for other reasons. The tidal forces of the SMBH disrupt a binary with a separation larger than the Hill-radius \citep{Binney_08} during pericenter passage. For a bound, equal mass binary there exists no value of the unknown line-of-sight distance $|z|$ that would make the binary survive a pericenter passage (figure~\ref{fig:z_Peri}). Also, 
the destruction timescale for such a soft binary in a distance of 3.5'' to the SMBH is only 19000 years (calculated for the observed projected separation of E2 and E4.0, \citet{Binney_08}). Furthermore, a mass of $180\,M_{\odot}$ per star seems unrealistic, but  this mass estimate could be too high due to the measurement errors. A mass per star of $85\,M_\odot$ is excluded only at the $2\sigma$-level.
Taken together, these arguments make the binary hypothesis unlikely, and it seems more natural to assume that
additional mass is needed for binding E2 and E4.0.

\subsection{Acceleration limits make an IMBH less likely}
\label{acclimit}
A binding IMBH can manifest itself in the accelerations it imposes on surrounding stars. Vice versa, the absence of accelerations
makes it more likely that IRS13E is a chance occurrence. Thus, we calculate now the probability for observing the observed (insignificant) accelerations, firstly for the case that a binding IMBH is present and secondly for the case that no additional mass is present in IRS13E. 

\subsubsection{Binding IMBH present}
 From our data we can only test for accelerations in the proper motions:
\begin {equation}
 a_\mathrm{2D}=\frac {r_\mathrm{2D}}{R_\mathrm{3D}} \cdot \frac {G\, M_\mathrm{IMBH}}{R_\mathrm{3D}^2}=G\, M_\mathrm{IMBH}\,\frac {r_\mathrm{2D}}{(r_\mathrm{2D}^2+z^2)^{3/2}}
\label{a2d}
\end {equation}

We do not measure any significant accelerations (table~\ref{tab:one_motion}). Firstly, this sets only limits on the location of the stars relative to the IMBH along the line of sight (z).
But secondly, it is unlikely that a cluster extends much more in the z-dimension than in the plane of sky, and this actually constrains the mass $M_\mathrm{IMBH}$ of the IMBH.

We use again as in section  section~\ref{clustermass} the Monte Carlo simulated cluster, which are similar to IRS13E. For all this clusters, we calculate the expected values of $a_\mathrm{2D}$. 

\begin{figure}
\begin{center}
\includegraphics[width=0.72\columnwidth,angle=-90]{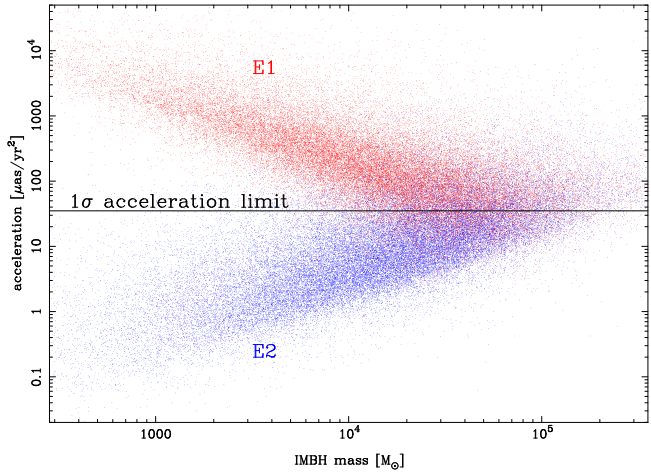}
\caption{Plane of sky accelerations of the simulated stars in our Monte Carlo clusters. 
Every red dot marks an acceleration of E1 in one of the 50000 simulated clusters. The blue dots stands for E2 in the same simulations. The horizontal line represents our acceleration limit. The region above the line is excluded at the $1\sigma$-level.  
} \label{fig:_acc_E12}
\end{center}
\end{figure}

Figure~\ref{fig:_acc_E12} shows that E2 is nearly not constraining, since for nearly all simulated clusters the expected value of $a_\mathrm{2D}$ is smaller than the $1\sigma$ acceleration limit. In case of E4.0 the expected accelerations are similar to E2. Since the measurement errors of E4.0 are four times larger than for E1 and E2, this star is of least importance. As consequence, only the acceleration limit of E1 is important. This is due to the fact that its velocity differs from the one of E2 and E4.0. As pointed out by \citet{Schoedel_05} in their figure 2, the IMBH is likely closest to E1. Therefore its motion is most sensitive to the mass of the IMBH.

It is apparent in figure~\ref{fig:_acc_E12} that E1 excludes an IMBH of lower mass more efficiently than an IMBH of higher mass.
This seems paradoxical, but results from the fact that the lower the mass is, the more likely it is that the IMBH resides close to E1.
If the IMBH is close to E1, the full velocity difference between E1 and the other stars is due to E1 and one can write.
\begin{equation}
\frac{m_\mathrm{E1} (\delta v)^2}{r_\mathrm{E1}}=\frac {G \,M_\mathrm{IMBH}\, m_\mathrm{E1}}{r_\mathrm{E1}^2}
\rightarrow
r_\mathrm{E1} = \frac{M_\mathrm{IMBH}}{(\delta v)^2}
\end{equation}
Reducing $M_\mathrm{IMBH}$ thus lowers $r_\mathrm{E1}$ if the velocity difference has to stay at the measured value. For the acceleration $a_\mathrm{E1}$ follows
\begin{equation}
a_\mathrm{E1}\propto \frac {M_\mathrm{IMBH}}{r_\mathrm{E1}^2} \propto \frac{(\delta v)^4}{M_\mathrm{IMBH}}
\end{equation}
This anti-proportionality of acceleration to $M_\mathrm{IMBH}$ is apparent in the simulations (red points in figure~\ref{fig:_acc_E12}).

We now calculate how probable the observed accelerations are, if one assumes a certain $M_\mathrm{IMBH}$. Hence, we
use those simulated clusters from section~\ref{clustermass} that fall into the corresponding mass bin and 
calculate for each case the  $\chi^2$ for the observed accelerations given their errors and the simulated acceleration values.
This in turn is then converted to a probability.
Because all simulated clusters are equally likely, we use the average over this mass bin as probability for obtaining the measured accelerations (figure~\ref{fig:_tot_prob}).  

All IMBHs with $M_\mathrm{IMBH}<10000\,M_{\odot}$ have a probability of less than 0.9\% to be consistent with our acceleration limits. At the most probable mass of $14000\,M_{\odot}$ the probability of not detecting an acceleration is 2.5\%.
For all mass bins the probability of $M_\mathrm{IMBH}>0$ is less than 3\%. The mass that yields acceleration values that are least unlikely is about 60000 M$_{\odot}$.

The main uncertainty for the derived probabilities is due to the assumption on the semi-major axis distribution $a=0.4'' \times p^{0.9}$. Using $0.28''$ or $0.52''$ instead of $0.4''$ (which makes the simulated clusters still look reasonably similar to IRS13E) changes the resulting probabilities by up to a factor of $\approx 2$. 

\subsubsection{No additional mass}
If there is no IMBH in IR13E we expect zero acceleration. We calculate the $\chi^2$ to the accelerations for this case, too, and obtain a probability of 39 \% for observing the measured acceleration values under the assumption that no additional mass is present in IRS13E.

The mass estimate of $\approx 1000\,M_\odot$ from the assumption that only E2 and E4 are bound would yield an acceleration of $\approx3\,\mu$as/yr$^2$. This is much smaller than our measurement accuracy and we cannot test this hypothesis.

\subsection{Proper motion of Sgr~A*}
\label{sec:radio}

\subsubsection{Binding IMBH present}
 \citet{Gualandris_09} show that the proper motion of Sgr~A* perpendicular to the plane of the Galaxy \citep{Reid_04} is the most promising method for excluding an IMBH at distances similar to that of IRS13E. For IRS13E itself, we can take advantage of the fact that
we know the velocity of the potential cluster and hence of the proposed IMBH. Thus we can predict the expected reflex motion of Sgr~A* via momentum conservation. We do not need to assume circular motion in the plane of sky, as \citet{Gualandris_09} had to.

We use the value from \citet{Reid_04} who find that the motion perpendicular to the Galactic plane is $-0.4 \pm 0.9\,$km/s.
We estimate the uncertainty of the motion of the IMBH simply by the velocity dispersion of the potential cluster IRS13E. 

We proceed as in section~\ref{acclimit}. For a given mass bin, we calculate the expected reflex motion of Sgr~A*
due to the IMBH the motion of which is varied by the estimated uncertainty. We compare the such calculated reflex motion with the measured value. The corresponding probabilities we again obtain from the $\chi^2$ values. Finally we use the average per mass bin to estimate the probability of the radio data of Sgr~A* under the assumption that an IMBH of that mass is present (figure~\ref{fig:_tot_prob}). The small insignificant proper motion of Sgr~A* is quite probable
for small ($<20000 M_{\odot}$) IMBH masses. But for $M_\mathrm{IMBH}>46000 M_{\odot}$ the data are inconsistent with the model at the 90\% level.

\subsubsection{No additional mass}
We also calculate probability for that the measured motion occurs when no additional mass is present. We obtain a probability of 65.7\%.

\subsection{Total probabilities}
\label{totprob}

\subsubsection{Binding IMBH present}

\begin{figure}
\begin{center}
\includegraphics[width=0.99\columnwidth]{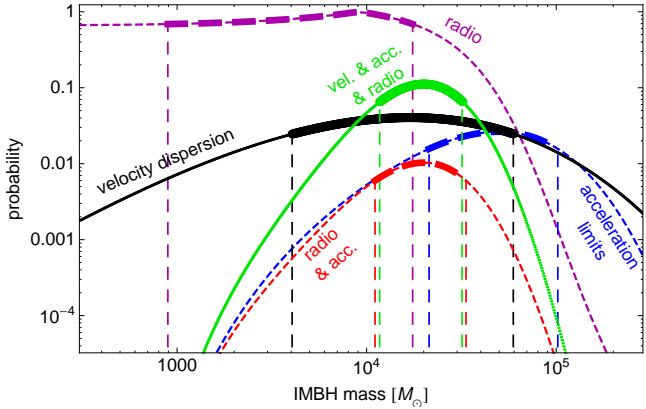}
\caption{Probabilities for different IMBH masses, if IRS13E is a gravitationally bound star cluster of 3 stars (E1,E2, E4.0).
Solid lines are probability densities per logarithmic mass bin of size $\Delta M/M=0.15$ and are normalized to 1. Dashed curves are probabilities calculated binwise. The thick regions mark the respective $68.3$ \% confidence regions of the most likely masses.
Black, solid: probability distribution of the mass of the IMBH from the measured velocity differences.
Blue, dashed: probability to observe the measured acceleration limits for an assumed mass of the IMBH.
Magenta, dashed: probability to observe measured radio motion of Sgr~A* for an assumed mass of the IMBH. 
Red, dashed: combining acceleration and radio information yields for every mass bin the probability that the observed values for the accelerations and radio motions occur. 
Green, solid: Probability distribution of the mass of the IMBH, combining velocity, acceleration and radio-based information,
}\label{fig:_tot_prob}
\end{center}
\end{figure}

In section~\ref{three} we have shown that if an IMBH massive enough to bind IRS13E is present, its most likely mass is (given the velocity differences between E1, E2 and E4)
\begin{equation}
M_\mathrm{IMBH}^\mathrm{pos,\,vel}=17000^{+43000}_{-13000}\,M_{\odot}.
\end{equation}
The corresponding distribution of probabilities $p_\mathrm{pos,\,vel} (m)$ is the solid, black curve in figure~\ref{fig:_tot_prob}, which is normalized to 1, since it assumes the presence of an IMBH.

Next, we calculate combined probabilities. For each mass bin $m$, the probability that the measured acceleration limits and the radio data are compatible with an IMBH of mass $m$ is
\begin{equation}
p_\mathrm{acc,\,radio}(m)= p_\mathrm{acc}(m) \times p_\mathrm{radio}(m)
\end{equation}
(figure~\ref{fig:_tot_prob}; red, dashed curve). Hence, at any given mass the probability for the data under the assumption that an IMBH is present is smaller than 0.8\%. 

Further, we can calculate the most likely probability value, by evaluating a weighted average of $p_\mathrm{acc,\,radio}(m)$ over all mass bins, using as weights $p_\mathrm{pos,\,vel} (m)$, i.e.
\begin{equation}
\label{eqcomb}
\bar{p}_\mathrm{comb} = \frac{\int p_\mathrm{acc,\,radio}(m)\times p_\mathrm{pos,\,vel} (m)\, dm}{ \int p_\mathrm{pos,\,vel} (m)\, dm},
\end{equation}
where the denominator equals 1 due to our normalization. The result is 
\begin{equation}
\bar{p}_\mathrm{comb} = 0.4\%.
\end{equation}
This value is uncertain by a factor of $\approx 2$.

Normalizing the nominator from equation~\ref{eqcomb} to 1, i.e. 
\begin{equation}
p_\mathrm{comb}(m) = \frac{p_\mathrm{acc,\,radio}(m)\times p_\mathrm{pos,\,vel} (m)\,}{\int p_\mathrm{acc,\,radio}(m)\times  p_\mathrm{pos,\,vel} (m)\, dm},
\end{equation}
is equivalent to a Bayesian approach, for obtaining a combined probability distribution of the mass. This results in the green, solid curve in figure~\ref{fig:_tot_prob}. It is normalized to 1 by construction which reflects the fact that the presence of an IMBH is assumed.
The most likely mass of an IMBH is thus 
$$M_\mathrm{IMBH}^\mathrm{comb}=20000^{+12000}_{-8000} \,M_{\odot}.$$

\subsubsection{No additional mass}

For the case that no additional mass is present, we obtain $p_\mathrm{acc,\,radio}(m=0)= 26$\%.
In section ~\ref{3_stars} we showed that
$p_\mathrm{pos,vel} (m=0)$ is 0.8\%, which yields $p_\mathrm{comb}(m=0) \approx 0.2$\%. The error on this value is a factor of $\approx 3$.

\section{Discussion}
\label{sec:discussion}

We are left with a puzzle. The probability to obtain the observed data (positions, velocities, accelerations, radio reflex motion of Sgr~A*) when no IMBH is present is 0.2\%. Similarly unlikely is the probability of getting the same data if one assumes the presence of an IMBH. For any mass assumed, the probability is smaller than 0.8\%. Hence, both scenarios are similarly unlikely given their errors of about a factor 2.5.
We therefore consider additional arguments.

\subsection{Dynamics of the young, massive stars}
The clockwise disk of early-type stars in the GC is significantly warped \citep{Bartko_09}. Such a warp can be the consequence of another non-spherical mass \citep{Nayakshin_05,Loeckmann_09}, and thus an IMBH might manifest itself via the warp. However, there are many other possibilities to explain the warp: The counterclockwise disk might suffice, or the warp can also happen in the gaseous phase during the disk formation \citep{Hobbs_09}. Dynamical friction will make an IMBH spiral down to the SMBH \citep{Fujii_09}, such that it would stay only for a very short time in the radial region of the disks. Hence, an IMBH is a less efficient warper per mass than a disk.

One possible origin of a potential IMBH is the core collapse of a young, massive star cluster \citep{Zwart_02}. IRS13E could be the remainder of such an object that currently is spiraling in. \citet{Schoedel_05} firstly indicated that using the simulations of \citet{Zwart_02} a cluster of an unrealistic high mass is necessary for a such massive IMBH as in IRS13E.  In addition, we compare the number of young, massive stars in the GC with the expection for the most likely mass of an IMBH of 
 $20000\,M_\odot$.
 
\begin{figure}
\begin{center}
\includegraphics[width=0.7\columnwidth,angle=-90]{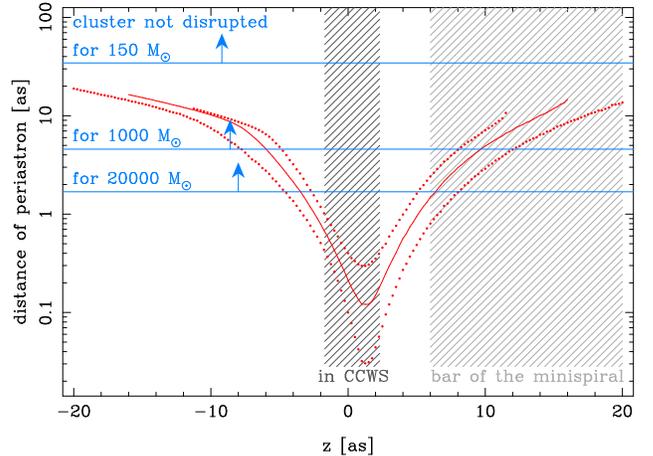}
\caption{Periastron distance of IRS13E for possible locations along the line of sight (z) in the point mass potential of the SMBH. The dotted lines mark the $1\sigma$-error band. Considering tidal disruption, only the $z$ positions corresponding to periastron distances above the blue lines are possible for the respective masses. The two lower lines for 1000 and $20000\,M_\odot$ assume a cluster with an IMBH, the upper line is for the case that E2 and E4.0 form a binary with $150\,M_\odot$ per star.
The shaded region close to $z=0''$ corresponds to the cases in which IRS13E is in the counterclockwise disk (CCWS) \citep{Bartko_09}. The shaded region to the right corresponds to a location of IRS13E in the minispiral \citep{Paumard_04}, which is possible for all configurations. It is not possible, that an IMBH binds IRS13E, and that at the same time IRS13E is located in counterclockwise disk. 
} \label{fig:z_Peri}
\end{center}
\end{figure}

In the simulations of \citet{Zwart_02, Zwart_04, Fujii_09} the IMBH obtains a mass of 0.1\% to 9\% of the total initial mass of the cluster, depending on the assumptions on the IMF, the radii and winds of very massive stars, the density of the cluster and whether a single supernova stops the formation of the IMBH. Thus, for the most likely IMBH mass one needs an initial cluster mass of at least $2\times10^5\,M_\odot$ \citep{Fujii_09}. 
Assuming a Salpeter IMF between 1 and $140\,M_{\odot}$ \citep{Fujii_09}, stars more massive than $40\,M_{\odot}$ would merge into the IMBH. The lighter stars should still be observable today, and one would expect to find 790 O-stars ($20-40\,M_\odot$) and 1950 early B-stars ($10-20\,M_\odot$).

In the GC there are in total only $\approx290$ O/WR-stars, correcting the currently detected number of such stars upward by a factor of 2.4, as estimated from incomplete areal coverage and detection incompleteness (figure~2 in \citet {Bartko_09b}). However, not all of them are compatible with the idea that they originate from IRS13E. 

Figure~\ref{fig:z_Peri} shows that it is not possible that an IMBH binds IRS13E, and that at the same time IRS13E is located in the counterclockwise disk. Hence, the inspiraling cluster scenario requires that the orbits of the stars stripped off the cluster are found in edge-on orbits in a common orbital plane. We estimate that at most 50 O/WR-stars could have such orbits. We observe, however, a factor 15 fewer O-stars than expected for an IMBH of $20000\,M_\odot$. For the bright B-stars, the discrepancy is even larger. We conclude that the formation of an IMBH of $20000\,M_{\odot}$ in IRS13E is inconsistent with the observed number of young stars.

\subsection{X-ray and radio emission}
Apart from Sgr~A*, IRS13E is the only
source with significant emission detectable in the NIR, radio, and
X-ray regime \citep{Zhao_98,Wang_06}.
At $13\,$mm, the morphology of the minispiral is similar to the one in L'-band. This emission traces hot gas, probably associated with the dust visible in the NIR.

The X-ray spectrum of IRS13E \citep{Wang_06} is similar to the diffuse X-ray emission from the local background and from the region around Sgr~A*. \citet {Coker_02} argue, that IRS13E is a wind nebula excited by a star, which left the Luminous Blue Variable (LBV) phase a short time ago. Maybe this scenario also works for the two WR-stars the winds of which probably collide in E3. The accuracy of the position of the X-ray source is not sufficient to decide whether the X-ray source can be identified with any of the bright NIR sources in IRS13E.

The observed flux in IRS13E is $L_\mathrm{X} (2-10\,\mathrm{keV})=2\times 10^{33}\,$erg/s \citep{Wang_06}, 
corresponding to $\approx 10^{-9}$ Eddington luminosities ($L_\mathrm{Edd}$) for an IMBH of $20000\,M_\odot$. This number is much smaller than for the X-ray binaries detected in the GC \citep{Muno_05}. The faintness of the X-ray source in IRS13E was  the main argument of \citet{Schoedel_05} against an IMBH in IRS13E.
However, Sgr~A* itself is radiating in quiescence with only  $L_\mathrm{X}(2-10\, \mathrm{keV})=4\times10^{-12}L_\mathrm{Edd}$ \citep{Baganoff_etal_2001_FirstXrayFlare} such that a low radiative efficiency does not exclude the presence of an IMBH right away. A comparison on basis of the X-ray flux only is difficult, since the environment and accretion state might differ significantly between Sgr~A* and an IMBH in IRS13E.

There is no variability reported for IRS13E in the X-rays \citep{Muno_05}, despite the fact that during roughly half of the observations of Sgr~A* IRS13E was observed, too. Many black holes are variable in the X-ray domain (examples are X-ray binaries \citep{Muno_05} and Sgr~A* \citep{Baganoff_etal_2001_FirstXrayFlare,Porquet_03}). In section~\ref{sec:struc} we argue that the dust seen in IRS13E is physically connected with the group of stars. Comparing the Br-$\gamma$ emission in IRS13E and close to Sgr~A* shows that the gas reservoir in the IRS13E is much richer than close to the SMBH. Taken together, this makes it unlikely  that the reported, steady source is due to an IMBH.

\section{Conclusions and summary}

We present a detailed analysis of the potential cluster IRS13E in the GC. We use AO-based images in H- and Ks-band from 2002 to 2009 for identifying objects in IRS13E and for performing astrometry. We detect 19 objects on most high Strehl ratio images in Ks-band and 15 of them also in H-band. The proper motions of the objects are well determined.

In addition, we characterize the SED of the objects in IRS13E using AO-based integral field spectroscopy (1.45 to $2.45\,\mu$m) for the brighter objects and (narrow-band) imaging from H- ($1.66\,\mu$m) to L'-band ($3.80\,\mu$m) for the fainter objects.
We fit the SED of the objects by a model consisting of two blackbodies and an extinction parameter. From this we conclude
that 13 objects are dust clumps without embedded stars and that the extinction appears to be constant in the IRS13E region.
Three objects are fainter stars: One is a spectroscopically identified late-type star. The two fainter stars of the magnitude of the red clump are consistent with the expected number of stars per area at this magnitude bin. Thus, these three stars are likely background stars. Therefore, IRS13E is only a concentration of the remaining three early-type stars E1, E2 and E4.0. 
 
They have the following properties:
\begin{itemize}
\item E2 and E4.0 are probable located close in 3D, because of the bright hot dust clumps between them.
\item A chance association of three early-type stars has a probability of around 0.2\%. Depending on the assumptions on dynamical properties of the stellar population this value varies between 0.06\% and 0.6\%.
\item 
In the case of a binding mass, we use Monte Carlo Simulations of a cluster of three stars around all reasonable masses of IMBHs (300 to $3\times 10^5 M_{\odot}$). From those simulations we select the ones that agree with the constraints from the observed positions and 
velocities  of the stars. The most likely mass is about $20000\,M_{\odot}$. This mass exceeds the stellar mass (inclusive faint, invisible stars) in IRS13E.
\item We find that at any given mass less than 0.8\% of the previously selected simulations agree with the observed acceleration limits of the three stars and the fact that Sgr~A* appears to be at rest. This value is uncertain by a factor of $\approx 2$.
\item There are roughly 15 times too few young stars in the GC for the formation of an IMBH as massive as $20000\,M_\odot$.
\item The weak and non-variable X-ray source in IRS13E makes the presence of an IMBH  unlikely.  
\end{itemize}

Overall, we conclude that it is more likely that IRS13E does not host an IMBH. 

\bibliographystyle{apj}
\bibliography{mspap}

\begin{thebibliography}{79}
\expandafter\ifx\csname natexlab\endcsname\relax\def\natexlab#1{#1}\fi

\bibitem[{{Allen} {et~al.}(1990){Allen}, {Hyland}, \& {Hillier}}]{Allen_90}
{Allen}, D.~A., {Hyland}, A.~R., \& {Hillier}, D.~J. 1990, \mnras, 244, 706

\bibitem[{{Baganoff} {et~al.}(2001){Baganoff}, {Bautz}, {Brandt}, {Chartas},
  {Feigelson}, {Garmire}, {Maeda}, {Morris}, {Ricker}, {Townsley}, \&
  {Walter}}]{Baganoff_etal_2001_FirstXrayFlare}
{Baganoff}, F.~K., {et~al.} 2001, \nat, 413, 45

\bibitem[{{Bartko} {et~al.}(2009){Bartko}, {Martins}, {Fritz}, {Genzel},
  {Levin}, {Perets}, {Paumard}, {Nayakshin}, {Gerhard}, {Alexander},
  {Dodds-Eden}, {Eisenhauer}, {Gillessen}, {Mascetti}, {Ott}, {Perrin},
  {Pfuhl}, {Reid}, {Rouan}, {Sternberg}, \& {Trippe}}]{Bartko_09}
{Bartko}, H., {et~al.} 2009, \apj, 697, 1741

\bibitem[{Bartko {et~al.}(2010)Bartko, Martins, Trippe, Fritz, Genzel, Ott,
  Eisenhauer, Gillessen, Paumard, Alexander, Dodds-Eden, Gerhard, Levin,
  Mascetti, Nayakshin, Perets, Perrin, Pfuhl, Reid, Rouan, Zilka, \&
  Sternberg}]{Bartko_09b}
Bartko, H., {et~al.} 2010, ApJ, 708, 834

\bibitem[{{Bik} {et~al.}(2006){Bik}, {Kaper}, \& {Waters}}]{Bik_06}
{Bik}, A., {Kaper}, L., \& {Waters}, L.~B.~F.~M. 2006, \aap, 455, 561

\bibitem[{{Binney} \& {Tremaine}(2008)}]{Binney_08}
{Binney}, J., \& {Tremaine}, S. 2008, {Galactic Dynamics: Second Edition}
  (Princeton University Press)

\bibitem[{{Blum} {et~al.}(1996){Blum}, {Sellgren}, \& {Depoy}}]{Blum_96}
{Blum}, R.~D., {Sellgren}, K., \& {Depoy}, D.~L. 1996, \apj, 470, 864

\bibitem[{{Bonnell} \& {Rice}(2008)}]{Bonnell_08}
{Bonnell}, I.~A., \& {Rice}, W.~K.~M. 2008, Science, 321, 1060

\bibitem[{{Bonnet} {et~al.}(2003){Bonnet}, {Str{\"o}bele}, {Biancat-Marchet},
  {Brynnel}, {Conzelmann}, {Delabre}, {Donaldson}, {Farinato}, {Fedrigo},
  {Hubin}, {Kasper}, \& {Kissler-Patig}}]{Bonnet_03}
{Bonnet}, H., {et~al.} 2003, in Society of Photo-Optical Instrumentation
  Engineers (SPIE) Conference Series, Vol. 4839, Society of Photo-Optical
  Instrumentation Engineers (SPIE) Conference Series, ed. {P.~L.~Wizinowich \&
  D.~Bonaccini}, 329--343

\bibitem[{{Buchholz} {et~al.}(2009){Buchholz}, {Sch{\"o}del}, \&
  {Eckart}}]{Buchholz_09}
{Buchholz}, R.~M., {Sch{\"o}del}, R., \& {Eckart}, A. 2009, \aap, 499, 483

\bibitem[{Christopher {et~al.}(2005)Christopher, Scoville, Stolovy, \&
  Yun}]{Christopher:2005p1402}
Christopher, M.~H., Scoville, N.~Z., Stolovy, S.~R., \& Yun, M.~S. 2005, ApJ,
  622, 346

\bibitem[{{Coker} {et~al.}(2002){Coker}, {Pittard}, \& {Kastner}}]{Coker_02}
{Coker}, R.~F., {Pittard}, J.~M., \& {Kastner}, J.~H. 2002, \aap, 383, 568

\bibitem[{{Crowther}(2007)}]{Crowther_07}
{Crowther}, P.~A. 2007, \araa, 45, 177

\bibitem[{{Diolaiti} {et~al.}(2000){Diolaiti}, {Bendinelli}, {Bonaccini},
  {Close}, {Currie}, \& {Parmeggiani}}]{Diolaiti_etal2000_StarFinder}
{Diolaiti}, E., {Bendinelli}, O., {Bonaccini}, D., {Close}, L.~M., {Currie},
  D.~G., \& {Parmeggiani}, G. 2000, in Presented at the Society of
  Photo-Optical Instrumentation Engineers (SPIE) Conference, Vol. 4007, Proc.
  Society of Photo-Optical Instrumentation Engineer Vol. 4007, p. 879-888,
  Adaptive Optical Systems Technology, Peter L. Wizinowich; Ed., ed. P.~L.
  {Wizinowich}, 879--888

\bibitem[{{Draine}(1989)}]{Draine_89}
{Draine}, B.~T. 1989, in ESA Special Publication, Vol. 290, Infrared
  Spectroscopy in Astronomy, ed. {E.~B{\"o}hm-Vitense}, 93--98

\bibitem[{Eckart {et~al.}(2004)Eckart, Moultaka, Viehmann, Straubmeier, \&
  Mouawad}]{Eckart:2004p1170}
Eckart, A., Moultaka, J., Viehmann, T., Straubmeier, C., \& Mouawad, N. 2004,
  ApJ, 602, 760

\bibitem[{{Eckart} {et~al.}(2004){Eckart}, {Moultaka}, {Viehmann},
  {Straubmeier}, \& {Mouawad}}]{Eckart_04}
{Eckart}, A., {Moultaka}, J., {Viehmann}, T., {Straubmeier}, C., \& {Mouawad},
  N. 2004, \apj, 602, 760

\bibitem[{{Eisenhauer} {et~al.}(2003){Eisenhauer}, {Abuter}, {Bickert},
  {Bianchet-Marchet}, {Bonnet}, {Brynnel}, {Conzelmann}, {Delabre},
  {Donaldson}, {Farinato}, {Fedrigo}, {Genzel}, {Hubin}, {Iserlohe}, {Kasper},
  {Kissler-Patig}, {Monnet}, {Roehrle}, {Scheiber}, {Stroebele}, {Tecza},
  {Thatte}, \& {Weisz}}]{Eisenhauer_etal2003}
{Eisenhauer}, F., {et~al.} 2003, SPIE, 4841, 1548

\bibitem[{{Forrest} {et~al.}(1987){Forrest}, {Shure}, {Pipher}, \&
  {Woodward}}]{Forrest_87}
{Forrest}, W.~J., {Shure}, M.~A., {Pipher}, J.~L., \& {Woodward}, C.~E. 1987,
  in American Institute of Physics Conference Series, Vol. 155, The Galactic
  Center, ed. {D.~C.~Backer}, 153--156

\bibitem[{Fritz {et~al.}(2010)Fritz, Gillessen, Trippe, Ott, Bartko, Pfuhl,
  Dodds-Eden, Davies, Eisenhauer, \& Genzel}]{Fritz_09}
Fritz, T.~K., {et~al.} 2010, MNRAS, 401, 1177

\bibitem[{{Fujii} {et~al.}(2009){Fujii}, {Iwasawa}, {Funato}, \&
  {Makino}}]{Fujii_09}
{Fujii}, M., {Iwasawa}, M., {Funato}, Y., \& {Makino}, J. 2009, \apj, 695, 1421

\bibitem[{{Genzel} {et~al.}(1996){Genzel}, {Thatte}, {Krabbe}, {Kroker}, \&
  {Tacconi-Garman}}]{Genzel_96}
{Genzel}, R., {Thatte}, N., {Krabbe}, A., {Kroker}, H., \& {Tacconi-Garman},
  L.~E. 1996, \apj, 472, 153

\bibitem[{{Genzel} {et~al.}(2003){Genzel}, {Sch{\"o}del}, {Ott}, {Eisenhauer},
  {Hofmann}, {Lehnert}, {Eckart}, {Alexander}, {Sternberg}, {Lenzen},
  {Cl{\'e}net}, {Lacombe}, {Rouan}, {Renzini}, \& {Tacconi-Garman}}]{Genzel_03}
{Genzel}, R., {et~al.} 2003, \apj, 594, 812

\bibitem[{{Gerhard}(2001)}]{Gerhard_01}
{Gerhard}, O. 2001, \apjl, 546, L39

\bibitem[{{Ghez} {et~al.}(2003){Ghez}, {Duch{\^e}ne}, {Matthews}, {Hornstein},
  {Tanner}, {Larkin}, {Morris}, {Becklin}, {Salim}, {Kremenek}, {Thompson},
  {Soifer}, {Neugebauer}, \& {McLean}}]{Ghez_03}
{Ghez}, A.~M., {et~al.} 2003, \apjl, 586, L127

\bibitem[{{Ghez} {et~al.}(2008){Ghez}, {Salim}, {Weinberg}, {Lu}, {Do}, {Dunn},
  {Matthews}, {Morris}, {Yelda}, {Becklin}, {Kremenek}, {Milosavljevic}, \&
  {Naiman}}]{Ghez_08}
---. 2008, \apj, 689, 1044

\bibitem[{Gillessen {et~al.}(2009)Gillessen, Eisenhauer, Trippe, Alexander,
  Genzel, Martins, \& Ott}]{Gillessen_09}
Gillessen, S., Eisenhauer, F., Trippe, S., Alexander, T., Genzel, R., Martins,
  F., \& Ott, T. 2009, ApJ, 692, 1075

\bibitem[{{Gosling} {et~al.}(2009){Gosling}, {Bandyopadhyay}, \&
  {Blundell}}]{Gosling_09}
{Gosling}, A.~J., {Bandyopadhyay}, R.~M., \& {Blundell}, K.~M. 2009, \mnras,
  394, 2247

\bibitem[{{Gualandris} \& {Merritt}(2009)}]{Gualandris_09}
{Gualandris}, A., \& {Merritt}, D. 2009, ArXiv e-prints

\bibitem[{{Hansen} \& {Milosavljevi{\'c}}(2003)}]{Hansen_03}
{Hansen}, B.~M.~S., \& {Milosavljevi{\'c}}, M. 2003, \apjl, 593, L77

\bibitem[{{Hillier} \& {Miller}(1998)}]{Hillier_98}
{Hillier}, D.~J., \& {Miller}, D.~L. 1998, \apj, 496, 407

\bibitem[{{Hobbs} \& {Nayakshin}(2009)}]{Hobbs_09}
{Hobbs}, A., \& {Nayakshin}, S. 2009, \mnras, 394, 191

\bibitem[{{Hummer} \& {Storey}(1987)}]{Hummer_87}
{Hummer}, D.~G., \& {Storey}, P.~J. 1987, \mnras, 224, 801

\bibitem[{{Kobulnicky} \& {Fryer}(2007)}]{Kobulnicky_07}
{Kobulnicky}, H.~A., \& {Fryer}, C.~L. 2007, \apj, 670, 747

\bibitem[{{Krabbe} {et~al.}(1991){Krabbe}, {Genzel}, {Drapatz}, \&
  {Rotaciuc}}]{Krabbe_91}
{Krabbe}, A., {Genzel}, R., {Drapatz}, S., \& {Rotaciuc}, V. 1991, \apjl, 382,
  L19

\bibitem[{{Lenzen} {et~al.}(2003){Lenzen}, {Hartung}, {Brandner}, {Finger},
  {Hubin}, {Lacombe}, {Lagrange}, {Lehnert}, {Moorwood}, \&
  {Mouillet}}]{Lenzen_etal2003_NACO}
{Lenzen}, R., {et~al.} 2003, in Presented at the Society of Photo-Optical
  Instrumentation Engineers (SPIE) Conference, Vol. 4841, Instrument Design and
  Performance for Optical/Infrared Ground-based Telescopes. Edited by Iye,
  Masanori; Moorwood, Alan F. M. Proceedings of the SPIE, Volume 4841, pp.
  944-952 (2003)., ed. M.~{Iye} \& A.~F.~M. {Moorwood}, 944--952

\bibitem[{{Liszt}(2003)}]{Liszt_03}
{Liszt}, H.~S. 2003, \aap, 408, 1009

\bibitem[{{Lo} \& {Claussen}(1983)}]{Lo_83}
{Lo}, K.~Y., \& {Claussen}, M.~J. 1983, \nat, 306, 647

\bibitem[{{L{\"o}ckmann} \& {Baumgardt}(2009)}]{Loeckmann_09}
{L{\"o}ckmann}, U., \& {Baumgardt}, H. 2009, \mnras, 394, 1841

\bibitem[{{Lu} {et~al.}(2005){Lu}, {Ghez}, {Hornstein}, {Morris}, \&
  {Becklin}}]{Lu_05}
{Lu}, J.~R., {Ghez}, A.~M., {Hornstein}, S.~D., {Morris}, M., \& {Becklin},
  E.~E. 2005, \apjl, 625, L51

\bibitem[{{Lu} {et~al.}(2009){Lu}, {Ghez}, {Hornstein}, {Morris}, {Becklin}, \&
  {Matthews}}]{Lu_09}
{Lu}, J.~R., {Ghez}, A.~M., {Hornstein}, S.~D., {Morris}, M.~R., {Becklin},
  E.~E., \& {Matthews}, K. 2009, \apj, 690, 1463

\bibitem[{{Lucy}(1974)}]{Lucy_74}
{Lucy}, L.~B. 1974, \aj, 79, 745

\bibitem[{{Lutz} {et~al.}(1993){Lutz}, {Krabbe}, \& {Genzel}}]{Lutz_93}
{Lutz}, D., {Krabbe}, A., \& {Genzel}, R. 1993, \apj, 418, 244

\bibitem[{{Lutz} {et~al.}(1996){Lutz}, {Feuchtgruber}, {Genzel}, {Kunze},
  {Rigopoulou}, {Spoon}, {Wright}, {Egami}, {Katterloher}, {Sturm},
  {Wieprecht}, {Sternberg}, {Moorwood}, \& {de Graauw}}]{Lutz_96}
{Lutz}, D., {et~al.} 1996, \aap, 315, L269

\bibitem[{{Maillard} {et~al.}(2004){Maillard}, {Paumard}, {Stolovy}, \&
  {Rigaut}}]{Maillard_04}
{Maillard}, J.~P., {Paumard}, T., {Stolovy}, S.~R., \& {Rigaut}, F. 2004, \aap,
  423, 155

\bibitem[{{Maness} {et~al.}(2007){Maness}, {Martins}, {Trippe}, {Genzel},
  {Graham}, {Sheehy}, {Salaris}, {Gillessen}, {Alexander}, {Paumard}, {Ott},
  {Abuter}, \& {Eisenhauer}}]{Maness_07}
{Maness}, H., {et~al.} 2007, \apj, 669, 1024

\bibitem[{{Mart{\'{\i}}n-Hern{\'a}ndez}
  {et~al.}(2008){Mart{\'{\i}}n-Hern{\'a}ndez}, {Bik}, {Puga}, {N{\"u}rnberger},
  \& {Bronfman}}]{Hernandez_08}
{Mart{\'{\i}}n-Hern{\'a}ndez}, N.~L., {Bik}, A., {Puga}, E., {N{\"u}rnberger},
  D.~E.~A., \& {Bronfman}, L. 2008, \aap, 489, 229

\bibitem[{{Martins} {et~al.}(2007){Martins}, {Genzel}, {Hillier}, {Eisenhauer},
  {Paumard}, {Gillessen}, {Ott}, \& {Trippe}}]{Martins_07}
{Martins}, F., {Genzel}, R., {Hillier}, D.~J., {Eisenhauer}, F., {Paumard}, T.,
  {Gillessen}, S., {Ott}, T., \& {Trippe}, S. 2007, \aap, 468, 233

\bibitem[{{Martins} {et~al.}(2009){Martins}, {Hillier}, {Bouret}, {Depagne},
  {Foellmi}, {Marchenko}, \& {Moffat}}]{Martins_09}
{Martins}, F., {Hillier}, D.~J., {Bouret}, J.~C., {Depagne}, E., {Foellmi}, C.,
  {Marchenko}, S., \& {Moffat}, A.~F. 2009, \aap, 495, 257

\bibitem[{{Monnier} {et~al.}(2002){Monnier}, {Tuthill}, \&
  {Danchi}}]{Monnier_02}
{Monnier}, J.~D., {Tuthill}, P.~G., \& {Danchi}, W.~C. 2002, \apjl, 567, L137

\bibitem[{Moultaka {et~al.}(2005)Moultaka, Eckart, Sch{\"o}del, Viehmann, \&
  Najarro}]{Moultaka:2005p2410}
Moultaka, J., Eckart, A., Sch{\"o}del, R., Viehmann, T., \& Najarro, F. 2005,
  A{\&}A, 443, 163

\bibitem[{{Muno} {et~al.}(2005){Muno}, {Pfahl}, {Baganoff}, {Brandt}, {Ghez},
  {Lu}, \& {Morris}}]{Muno_05}
{Muno}, M.~P., {Pfahl}, E., {Baganoff}, F.~K., {Brandt}, W.~N., {Ghez}, A.,
  {Lu}, J., \& {Morris}, M.~R. 2005, \apjl, 622, L113

\bibitem[{{Mu{\v z}i{\'c}} {et~al.}(2007){Mu{\v z}i{\'c}}, {Eckart},
  {Sch{\"o}del}, {Meyer}, \& {Zensus}}]{Muzic_07}
{Mu{\v z}i{\'c}}, K., {Eckart}, A., {Sch{\"o}del}, R., {Meyer}, L., \&
  {Zensus}, A. 2007, \aap, 469, 993

\bibitem[{{Mu{\v z}i{\'c}} {et~al.}(2008){Mu{\v z}i{\'c}}, {Sch{\"o}del},
  {Eckart}, {Meyer}, \& {Zensus}}]{Muzic_08}
{Mu{\v z}i{\'c}}, K., {Sch{\"o}del}, R., {Eckart}, A., {Meyer}, L., \&
  {Zensus}, A. 2008, \aap, 482, 173

\bibitem[{{Nayakshin}(2005)}]{Nayakshin_05}
{Nayakshin}, S. 2005, \mnras, 359, 545

\bibitem[{{Nishiyama} {et~al.}(2009){Nishiyama}, {Tamura}, {Hatano}, {Kato},
  {Tanab{\'e}}, {Sugitani}, \& {Nagata}}]{Nishiyama_09}
{Nishiyama}, S., {Tamura}, M., {Hatano}, H., {Kato}, D., {Tanab{\'e}}, T.,
  {Sugitani}, K., \& {Nagata}, T. 2009, \apj, 696, 1407

\bibitem[{{Ott}(2002)}]{Ott_02}
{Ott}, T. 2002, PhD thesis, Ludwig-Maximilians-Universitaet Muenchen

\bibitem[{{Paumard} {et~al.}(2004){Paumard}, {Maillard}, \&
  {Morris}}]{Paumard_04}
{Paumard}, T., {Maillard}, J.-P., \& {Morris}, M. 2004, \aap, 426, 81

\bibitem[{{Paumard} {et~al.}(2006){Paumard}, {Genzel}, {Martins}, {Nayakshin},
  {Beloborodov}, {Levin}, {Trippe}, {Eisenhauer}, {Ott}, {Gillessen}, {Abuter},
  {Cuadra}, {Alexander}, \& {Sternberg}}]{Paumard_06}
{Paumard}, T., {et~al.} 2006, \apj, 643, 1011

\bibitem[{{Porquet} {et~al.}(2003){Porquet}, {Predehl}, {Aschenbach}, {Grosso},
  {Goldwurm}, {Goldoni}, {Warwick}, \& {Decourchelle}}]{Porquet_03}
{Porquet}, D., {Predehl}, P., {Aschenbach}, B., {Grosso}, N., {Goldwurm}, A.,
  {Goldoni}, P., {Warwick}, R.~S., \& {Decourchelle}, A. 2003, \aap, 407, L17

\bibitem[{{Portegies Zwart} {et~al.}(2007){Portegies Zwart}, {Gaburov}, {Chen},
  \& {G{\"u}rkan}}]{Zwart_07}
{Portegies Zwart}, S., {Gaburov}, E., {Chen}, H.-C., \& {G{\"u}rkan}, M.~A.
  2007, \mnras, 378, L29

\bibitem[{{Portegies Zwart} {et~al.}(2004){Portegies Zwart}, {Baumgardt},
  {Hut}, {Makino}, \& {McMillan}}]{Zwart_04}
{Portegies Zwart}, S.~F., {Baumgardt}, H., {Hut}, P., {Makino}, J., \&
  {McMillan}, S.~L.~W. 2004, \nat, 428, 724

\bibitem[{{Portegies Zwart} \& {McMillan}(2002)}]{Zwart_02}
{Portegies Zwart}, S.~F., \& {McMillan}, S.~L.~W. 2002, \apj, 576, 899

\bibitem[{{Reid}(1993)}]{Reid_93}
{Reid}, M.~J. 1993, \araa, 31, 345

\bibitem[{{Reid} \& {Brunthaler}(2004)}]{Reid_04}
{Reid}, M.~J., \& {Brunthaler}, A. 2004, \apj, 616, 872

\bibitem[{{Reid} {et~al.}(2007){Reid}, {Menten}, {Trippe}, {Ott}, \&
  {Genzel}}]{Reid_07}
{Reid}, M.~J., {Menten}, K.~M., {Trippe}, S., {Ott}, T., \& {Genzel}, R. 2007,
  \apj, 659, 378

\bibitem[{{Rieke} \& {Lebofsky}(1985)}]{Rieke_85}
{Rieke}, G.~H., \& {Lebofsky}, M.~J. 1985, \apj, 288, 618

\bibitem[{{Rieke}(1999)}]{Rieke_99}
{Rieke}, M.~J. 1999, in Astronomical Society of the Pacific Conference Series,
  Vol. 186, The Central Parsecs of the Galaxy, ed. {H.~Falcke, A.~Cotera,
  W.~J.~Duschl, F.~Melia, \& M.~J.~Rieke}, 32--+

\bibitem[{{Rousset} {et~al.}(2003){Rousset}, {Lacombe}, {Puget}, {Hubin},
  {Gendron}, {Fusco}, {Arsenault}, {Charton}, {Feautrier}, {Gigan}, {Kern},
  {Lagrange}, {Madec}, {Mouillet}, {Rabaud}, {Rabou}, {Stadler}, \&
  {Zins}}]{Rousset_etal_2003_NACO}
{Rousset}, G., {et~al.} 2003, in Presented at the Society of Photo-Optical
  Instrumentation Engineers (SPIE) Conference, Vol. 4839, Adaptive Optical
  System Technologies II. Edited by Wizinowich, Peter L.; Bonaccini, Domenico.
  Proceedings of the SPIE, Volume 4839, pp. 140-149 (2003)., ed. P.~L.
  {Wizinowich} \& D.~{Bonaccini}, 140--149

\bibitem[{{Sch{\"o}del} {et~al.}(2005){Sch{\"o}del}, {Eckart}, {Iserlohe},
  {Genzel}, \& {Ott}}]{Schoedel_05}
{Sch{\"o}del}, R., {Eckart}, A., {Iserlohe}, C., {Genzel}, R., \& {Ott}, T.
  2005, \apjl, 625, L111

\bibitem[{{Sch{\"o}del} {et~al.}(2009){Sch{\"o}del}, {Merritt}, \&
  {Eckart}}]{Schoedel_09}
{Sch{\"o}del}, R., {Merritt}, D., \& {Eckart}, A. 2009, \aap, 502, 91

\bibitem[{{Sch{\"o}del} {et~al.}(2010){Sch{\"o}del}, {Najarro}, {Muzic}, \&
  {Eckart}}]{Schoedel_09b}
{Sch{\"o}del}, R., {Najarro}, F., {Muzic}, K., \& {Eckart}, A. 2010, \aap, 511,
  A18+

\bibitem[{{Sch{\"o}del} {et~al.}(2002){Sch{\"o}del}, {Ott}, {Genzel},
  {Hofmann}, {Lehnert}, {Eckart}, {Mouawad}, {Alexander}, {Reid}, {Lenzen},
  {Hartung}, {Lacombe}, {Rouan}, {Gendron}, {Rousset}, {Lagrange}, {Brandner},
  {Ageorges}, {Lidman}, {Moorwood}, {Spyromilio}, {Hubin}, \&
  {Menten}}]{Schoedel_02}
{Sch{\"o}del}, R., {et~al.} 2002, \nat, 419, 694

\bibitem[{{Stead} \& {Hoare}(2009)}]{Stead_09}
{Stead}, J.~J., \& {Hoare}, M.~G. 2009, \mnras, 400, 731

\bibitem[{{Trippe} {et~al.}(2009){Trippe}, {Davies}, {Eisenhauer}, {Foerster
  Schreiber}, {Fritz}, \& {Genzel}}]{Trippe_09}
{Trippe}, S., {Davies}, R., {Eisenhauer}, F., {Foerster Schreiber}, N.~M.,
  {Fritz}, T.~K., \& {Genzel}, R. 2009, ArXiv e-prints

\bibitem[{{Trippe} {et~al.}(2008){Trippe}, {Gillessen}, {Gerhard}, {Bartko},
  {Fritz}, {Maness}, {Eisenhauer}, {Martins}, {Ott}, {Dodds-Eden}, \&
  {Genzel}}]{Trippe_08}
{Trippe}, S., {et~al.} 2008, \aap, 492, 419

\bibitem[{{Wang} {et~al.}(2006){Wang}, {Lu}, \& {Gotthelf}}]{Wang_06}
{Wang}, Q.~D., {Lu}, F.~J., \& {Gotthelf}, E.~V. 2006, \mnras, 367, 937

\bibitem[{{Zhao} \& {Goss}(1998)}]{Zhao_98}
{Zhao}, J.-H., \& {Goss}, W.~M. 1998, \apjl, 499, L163+

\bibitem[{{Zhao} {et~al.}(2009){Zhao}, {Morris}, {Goss}, \& {An}}]{Zhao_09}
{Zhao}, J.-H., {Morris}, M.~R., {Goss}, W.~M., \& {An}, T. 2009, \apj, 699, 186

\end{thebibliography}

\appendix

\section{A: Extracting the spectral energy distributions}
\label{sec:ob_seds}

\subsection{Deconvolution of SINFONI data}
In order to distinguish between objects in the SINFONI data, we deconvolve the cube in small bandpasses, taking into account the variation of the PSF with wavelength. We extract line maps at three wavelengths at which single stars prominently dominate the respective images: at $1.49\,\mu$m the dominant star is E4, at $1.70\,\mu$m and $2.11\,\mu$m it is E2. We manually remove apparent artifacts and smooth the PSF wings azimuthally. We then spectrally inter- and extrapolate these single-wavelength PSFs over the whole wavelength range, either resampled to 24 spectral channels or at the original spectral sampling. We then used two sets of PSFs to perform  deconvolution of the cube at the different samplings. We use the deconvolution at the original sampling for extracting the spectra of the WR stars E2 and E4.0. The resampled deconvolution is used for E3.0 and 5.0.

 We also use the  PSFs interpolated at the original sampling for creating a subtracted cube  in which the four brightest stars (E1, E2, E4.0 and the bright late-type star north-west of IRS13E) are subtracted. For this purpose we fit in each spectral plane the stars with Gaussians to determine their flux as a function of wavelength. Then we subtract the correspondingly scaled, interpolated PSFs from the original data cube. We use the resulting cube for analyzing the spectral features of the gas and the fainter stars.

We flux-calibrate the SINFONI data by using the early-type star E1, in the spectrum of which we manually replace the (few) spectral lines by continuum. The photometric errors from the SINFONI data are estimated by varying the selection of source pixels. During the fit of the such obtained data with an SED model, we rescale them such that the reduced $\chi^2=1$.

\subsection{Photometry from NACO images}
In order to measure the SED of the objects fainter or more confused than E5.0, we determine fluxes from the deconvolved broad-band and narrow-band images. We measure the flux of each object at each available wavelength by manually selecting the respective object pixels in the frames. For the broad-band images we estimate the flux errors from the standard deviation of the measured fluxes at the different epochs. For  the narrow-band images we simply adopt the same flux errors as obtained from the broad-band images.

For the broad-band images in H and Ks, we calibrate the magnitudes following \citet{Maness_07}\footnote{The calibration of \citet{Schoedel_09b} is about 0.3 mag fainter in both bands.}. 
We derive the narrow-band calibration from the broad-band calibration by using the empirical extinction law 
\begin{equation}
A_\lambda=A_\mathrm{Ks}*{\lambda/\lambda_\mathrm{Ks}}^{\alpha}=A_\mathrm{Ks}*{\lambda/\lambda_\mathrm{Ks}}^{-1.75}
\label{extlaw}
\end{equation}
from \citet{Draine_89}. As calibrators we use eight isolated, bright early-types stars with known temperatures within a distance of less than 2.5'' to IRS13E. 
Since there is no suitable L'-band calibrator close to IRS13E, we use an indirect method by extrapolating the dereddened SED of the calibrators to L'-band and applying the extinction law from \citet{Lutz_96} longward of Brackett-$\beta$ and short-ward the power law of \citet{Draine_89}. This results in 
$A_\mathrm{Ks}/A_\mathrm{L'}=1.75$. 
The slope of the power law in equation~\ref{extlaw} is uncertain: $\alpha=-1.6$ \citep{Rieke_85, Rieke_99}, $\alpha=-2$ \citep{Nishiyama_09} and $\alpha=-2.21$ \citep{Schoedel_09b}. Recent works may indicate that
 the extinction law used here is outdated \citep{Stead_09, Gosling_09}.
But at our precision the value of $\alpha$ does not affect the SED fitting because the shape of the derived SED is relatively insensitive to the value of $\alpha$ adopted. Nevertheless it introduces an uncertainty to the extinction values. 
Thus, it is important for the absolute luminosities of the early-type stars. To estimate these uncertainties, we use both the originally adopted extinction and
the law from \citet{Schoedel_09b} plus the magnitudes of the calibration stars in H, Ks and L' published by \citet{Schoedel_09b}.  

For the SED fits, we calculate the effective wavelength of the bands for each object. First, we apply an extinction value of $A_\mathrm{Ks}=2.8$ together with the atmospheric and the filter transmission to the source spectrum (either a blackbody of the output from the atmosphere modeling). For the broad-band data it is necessary to perform  a second iteration using the resulting SED in order to improve  the effective wavelength. 

\begin{figure}
\begin{center}
\includegraphics[width=0.53\columnwidth]{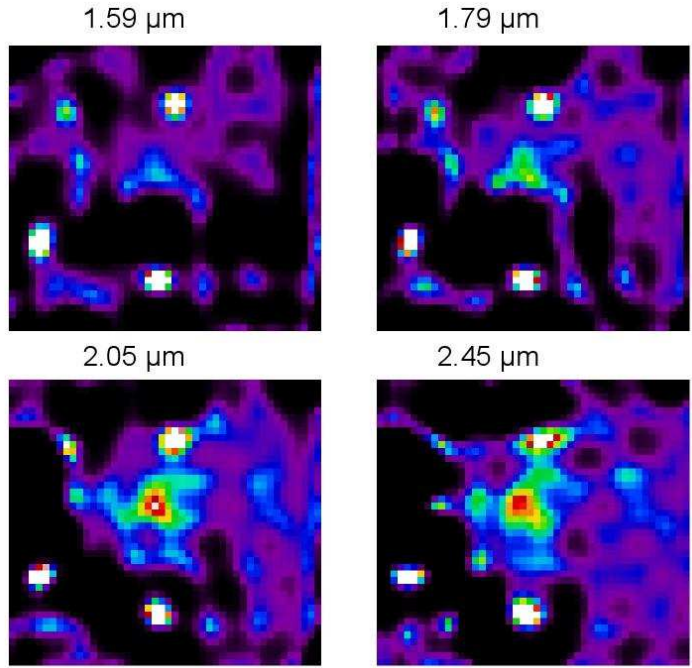}
\caption{Four of the 24 deconvolved narrowband images created from the high-quality H+K-band SINFONI cube of IRS13E. 
} \label{fig:cub_im}
\end{center}
\end{figure}

\section{B: Individual objects in IRS13E}
\label{sec:objects}

\subsection{IRS13E1}
E1 is a hot OB supergiant \citep{Paumard_06}. Its flux is constant in our data, and the spectrum does not reveal any signs of binarity. We use the H-band lines from the Brackett-series to determine the radial velocity (in K-band, the emission lines from nebular background are more dominant). We obtain $v_\mathrm{LSR}=134\pm10\,$km/s. This is $3\sigma$ different to $v_\mathrm{LSR}=71\pm20\,$km/s, the  Brackett-$\gamma$ based value reported in \citet{Paumard_06}. 

We compare our atmosphere model of E1 to a simple blackbody model of $T=26000 \pm 1000 \,$K and do not see differences between the two models which would matter for the broad-band photometry. 
Therefore, we fit the extinction using the photometric data and a blackbody model. We obtain $A_\mathrm{Ks}= 3.64 \pm 0.07$ (figure~\ref{fig:many_SED}). Contrary to \citet{Maillard_04}, we do not find a weak dust component.
This is likely because of the higher resolution of our data, since the SINFONI data show that the star is surrounded by gas probably also associated with dust. Fitting a supergiant model to the observed SED we obtain $\log L/L_{\odot} =6.23 \pm 0.07$. Thus E1 would be more luminous than all  known  supergiants of this spectral class. When using the calibration and extinction law of \citet{Schoedel_09b} we obtain   $\log L/L_{\odot} =5.90 \pm 0.07$, consistent with the brightest known OB supergiants.

\subsection{IRS13E2}
E2 is a Wolf-Rayet star of type WN8 \citep{Paumard_06, Martins_07}. There is no sign of binarity in its light curve or in its spectrum. 
The latter shows strong, broad emission lines. In addition, WR-stars have strong winds causing a near-infrared excess. Therefore it is necessary to use a full atmosphere model for the SED fit of E2. We manually adjust the temperature of a possible dust component and the extinction parameter to make the atmosphere model match the observed spectrum.

The SED fit yields a second blackbody component of around $T=712\,$K and $A_\mathrm{Ks}= 3.51$. This result is qualitatively consistent with the findings of \citet{Maillard_04}. The smaller dust temperature they find is likely caused by their assumption that E2 can be modeled with a blackbody. We find that the dust component is significant in L'-band, but in Ks-band the stellar component dominates (see Figure~\ref{fig:spec_E2}), contrary to the proposal by \citet{Martins_07}.  

We obtain for E2 $\log L/L_{\odot}=6.14\pm 0.1$ when using the standard calibration and extinction law of this paper and $\log L/L_{\odot}=5.81\pm 0.1$ when using calibration and extinction law of \citet{Schoedel_09b}. Thus. E2 is relatively luminous for its classification as WN8 star. 
E2 could be more massive and even younger ($\approx3\,$Myr) than most of the young, massive stars in the central parsec. However, due to the particular shape of the evolutionary tracks, E2 could equally well have the same age as the other stars of similar type. 
In their figure~23 \citet{Martins_07} gave for E2 only an upper limit in luminosity because of the unknown dust contribution at the time.
From our SED fit we were able to show that the Ks-luminosity is to about 96\% of stellar origin.
Therefore the luminosity of E2 in \citet{Martins_07}  is nearly identical with the value derived here.
In any case, E2 is brighter than the other WN8 stars in the GC. This fact is is unaffected by calibration uncertainties. Also, binarity would not be able to fully explain the luminosity of E2.

We use the $2.24\,\mu$m NIII line from the full atmosphere model for deriving the radial velocity of E2, which is more accurate compared to previous works.
We obtain $v_\mathrm{LSR}=65\pm30\,$km/s while \citet{Paumard_06} reported $v_\mathrm{LSR}=40\pm40\,$km/s and \citet{Maillard_04} gave $v_\mathrm{LSR}=30\,$km/s. All these values are consistent with each other.

\subsection{IRS13E3.1 to IRS13E3.4}
 The extracted PSFs of the SINFONI data are not good enough to measure the H-band part of the SEDs of the fainter objects around E3.0. Qualitatively, the spectra of E3.1 to E3.4 do not show any new spectral features compared to E3.0. There is only a gradual change in the spectral properties, such as the line velocities. Hence, these sources also likely consist mostly of gas and dust. This is confirmed by the SED fits, which
are based on the narrow-band images. Because of the faintness of the sources we cannot reliably fit the extinction, which we 
restrict for pure dust sources to $A_\mathrm{Ks}=3.4-3.9$. 

On most images in K-band (including the narrow-band images before 2005) and all L'-band data, the sources E3.1 and E3.2 are strongly overlapping. Therefore we fit a combined SED for the two (figure~\ref{fig:many_SED}). 
The properties of the fitted blackbody are similar to E3.0 apart from its brightness. A stellar component is not necessary, the $1\sigma$ limit is m$_\mathrm{Ks}=$18.0, and thus again we can exclude that a WR star of the most common types is hidden in E3.1 and E3.2. We suspect that \citet{Maillard_04} found a stellar component in the eastern part of their source E3 due to the same bias as for E3.0 (section~\ref{sece3}). Also for the less well constrained objects E3.3 and E3.4 (table~\ref{tab:spec_properties}) no stellar components are required. This conclusion does not depend on the assumption for the extinction.

\subsection{IRS13E3.5}
The SED of E3.5, derived photometrically, differs from the other sources in E3 (figure~\ref{fig:many_SED}). Its color is consistent with that of a star, and no dust is necessary to explain the SED of this object. 
Unfortunately, it is too confused and too faint for deriving its spectral type.
Any temperature between $4000\,$K and $19000\,$K is possible, which also implies an uncertainty of $\Delta A_\mathrm{Ks}=0.4$.
The absolute magnitude of E3.5 corresponds to the red clump, which makes it likely that E3.5 is an old, late-type star. In addition, this star does not share the proper motion of the three brightest stars in IRS13E. We conclude
that E3.5 most likely does not belong to IRS13E.

\subsection{IRS13E4.0}\label{sec_e4}
E4.0 is a Wolf-Rayet (WC9) star \citep{Paumard_06, Martins_07}.
Its radial velocity is very uncertain.  \citet {Paumard_06} report $v_\mathrm{LSR}=56 \pm 70\,$km/s,  \citet{Maillard_04} claim $v_\mathrm{LSR}=-30\,$km/s. From our atmosphere model (figure~\ref{fig:spec_E2}) including wind induced line broadening we obtain $v_\mathrm{LSR}=200 \pm 200\,$km/s.

As for E2 we manually adjust the atmosphere model with an additional dust component to match the measured spectrum (figure~\ref{fig:spec_E2}).  
Clearly two components are needed to describe the SED, but the derived values are very uncertain. 
The dust component with $T=1400\,$K is significant. It contributes 38\% of the flux in the Ks-band, as indicated by \citet{Martins_07}. Our measurement is more precise than the one from \citet{Maillard_04}, because we use a full atmosphere model and extend the wavelength coverage to L'-band. We obtain the same extinction as for E1.
The luminosity of E4.0 is  $\log L/L_\odot=5.23$. 
Because of the difficult modeling of this star and WC9 stars in general it is not possible to conclude whether E4.0 has an age similar to the other WC9 in the GC. 

\subsection{IRS13E4.1 to IRS13E4.3}
Besides  H- and Ks-band, E4.1 is detected also in L'-band and has overall a red SED.
It is too close to E4.0 to allow the broadband SED to be extracted from the SINFONI data.
The spectrum shows only gas lines and no stellar features.
For fitting the SED it was necessary to fix the extinction at the upper end of the possible range to $A_\mathrm{Ks}=3.9$. 
With this the data are fit well by a single blackbody of $T=844\pm 24\,$K.  
Any stellar component has to be fainter than $m_{\text{Ks}}=18.4$. 

E4.2 does not show any spectral features and is not detected in L'-band. 
Therefore, there  is less data than for the previous sources. We fix $A_\mathrm{Ks}=3.9$. From a single blackbody fit we get a temperature of less than $2800\,$K, which is too cold for a star of $m_{\text{Ks}}=15.4$. Thus, we need to add a second blackbody to the fit, which means we classify E4.2 as a star. 
As in the case for E3.5, this faint star matches the typical luminosity of red clump stars. Hence, it is most likely an old late-type star. Its motion, being slower westward than for the three bright stars in IRS13E,  argues against E4.2 being a physical member of  IRS13E.

For E4.3 we are only able to photometry in Ks-band, yielding $m_\mathrm{Ks}=15.5\pm0.1$.

\subsection{IRS13E5.0}
The source E5.0 is isolated enough to extract its broadband SED in the Ks-band from the SINFONI data (figure~\ref{fig:many_SED}), but because of its location close to the edge of the data cube, the SINFONI H-band measurements are not reliable. Hence, we combine the NACO-based fluxes with the SINFONI data. We allow for a global scaling factor to match the two data sets, which yields an offset of $0.24\,$mag. In any case, the fit does not change significantly by either including or excluding the SINFONI data.
The preferred fit yields $A_\mathrm{Ks}=3.22$ 
and a dusty blackbody. Thus, we fix the extinction value to the lowest value from the range assumed here to $A_\mathrm{Ks}=3.4$. Again, no stellar component is needed, and any star inside is fainter than $m_\mathrm{Ks}=18.5$.
The dust of E5.0 is with $T=700 \pm 8\,$K colder than for most other objects in E3 and E4.
\citet{Maillard_04} were only able to identify one source in E5. The scatter in their data points is so high that it is not clear how significant the stellar component is in their fits. 

\subsection{IRS13E5.1 to IRS13E5.4}

The other sources of E5 are also red but fainter than E5.0. They appear embedded in the brightest part of the minispiral.
It is not possible to measure their SEDs separated from the minispiral at our resolution, because the background is inhomogeneous and therefore the background subtraction too uncertain.
 This holds both for the NACO photometry and for the SINFONI data. In addition, the sources E5.2 and E5.3 are located on the deconvolution ring of the bright source E1. Thus, the fluxes of these two sources are biased to  high values, especially in H-band. Finally many of the objects are not detected in all our NACO bands. Consequently, the characterization of these source is considerably more uncertain than for the other sources.
The spectra of these sources show only gaseous emission lines. We again assume that the continuum emission is mainly caused by dust, and thus we restrict $A_\mathrm{Ks}$ to the range $3.4 - 3.9$.

The source E5.1 is a double source on the higher quality Ks-band images obtained since 2008, but we are not yet able to disentangle them.
This source is also detected in the L'-band. With our assumption for the extinction, the best fit is a blackbody with $T=947\pm25\,$K; any stellar component needs to be fainter than $m_{\text{Ks}}=17.8$.

The source E5.2 is not detected in H-band but in L'-band, where it is nearly overlapping with E5.3. We nevertheless simply assume that all of the L'-flux is caused by E5.2. With this we obtain a temperature of $T=642 \pm 70\,$K from the SED fit, which is consistent with the non-detection in the H-band.
Any  star hidden in E5.2 must be fainter than $m_\mathrm{Ks}=15.6$. Given the lower temperature, E5.2 is  likely a part of the minispiral. 

Contrary to E5.2, the object E5.3 is not detected in L'-band but in H-band. Our fit does not require a stellar component. The $1\sigma$-limit is $m_\mathrm{Ks}=16.0$ and we obtain a temperature of $T=978 \pm 524\,$K. The large error is caused by the smaller wavelength range on which the SED fit is based.

The source E5.4 is located close to E5.0. It is detected only in Ks- and the narrow bands. Hence we fit only a single blackbody with the restricted extinction interval. We obtain a temperature of $T=817 \pm 474\,$K. 

\subsection{IRS13E6}
\label{sec_E6}

\begin{figure}
\begin{center}
\includegraphics[width=0.29\columnwidth,angle=-90]{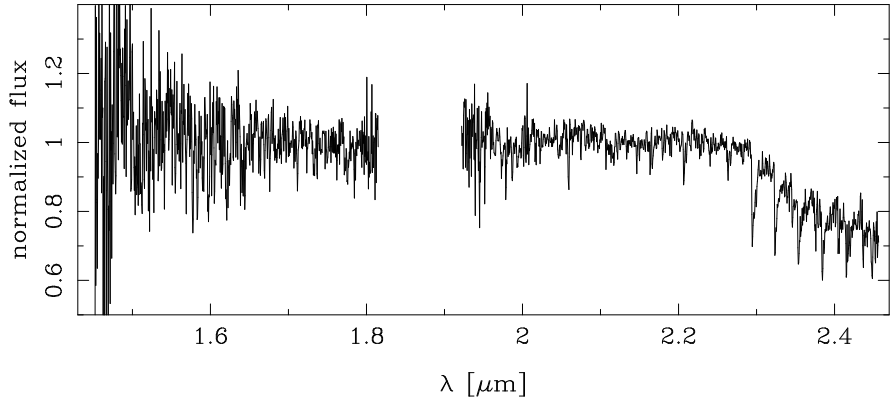}
\caption{SINFONI H+K-band spectrum of E6. 
} \label{fig:spec_E6}
\end{center}
\end{figure}

The spectrum (figure~\ref{fig:spec_E6}) of E6 ($m_{\text{Ks}}=13.82\pm 0.10$) shows CO-band heads in H- and Ks-band and lines of NaI and CaI. Thus, E6 is a late-type star with a spectral class around K3III. 
\citet{Maillard_04}  identified E6 as an O-star, but they used data from three relatively broad bands for their photometric identification. \citet{Buchholz_09} used narrow bands around the CO-band heads and determined correctly that E6 is a late-type star. We derive a radial velocity of $v_\mathrm{LSR}=106 \pm 10\,$km/s from the CO-band heads in Ks-band.  

In order to avoid biases, we exclude from the SED fit the bands which contain flux of the CO band heads.
The fitted extinction value of $A_{\text{Ks}}=3.68 \pm 0.09$ is consistent with the extinction of E1.
Due to its faintness E6 cannot be a red supergiant and  it is thus older than the three bright early-type stars.

\section{C: Astrometric data} 
\begin{figure}[h]
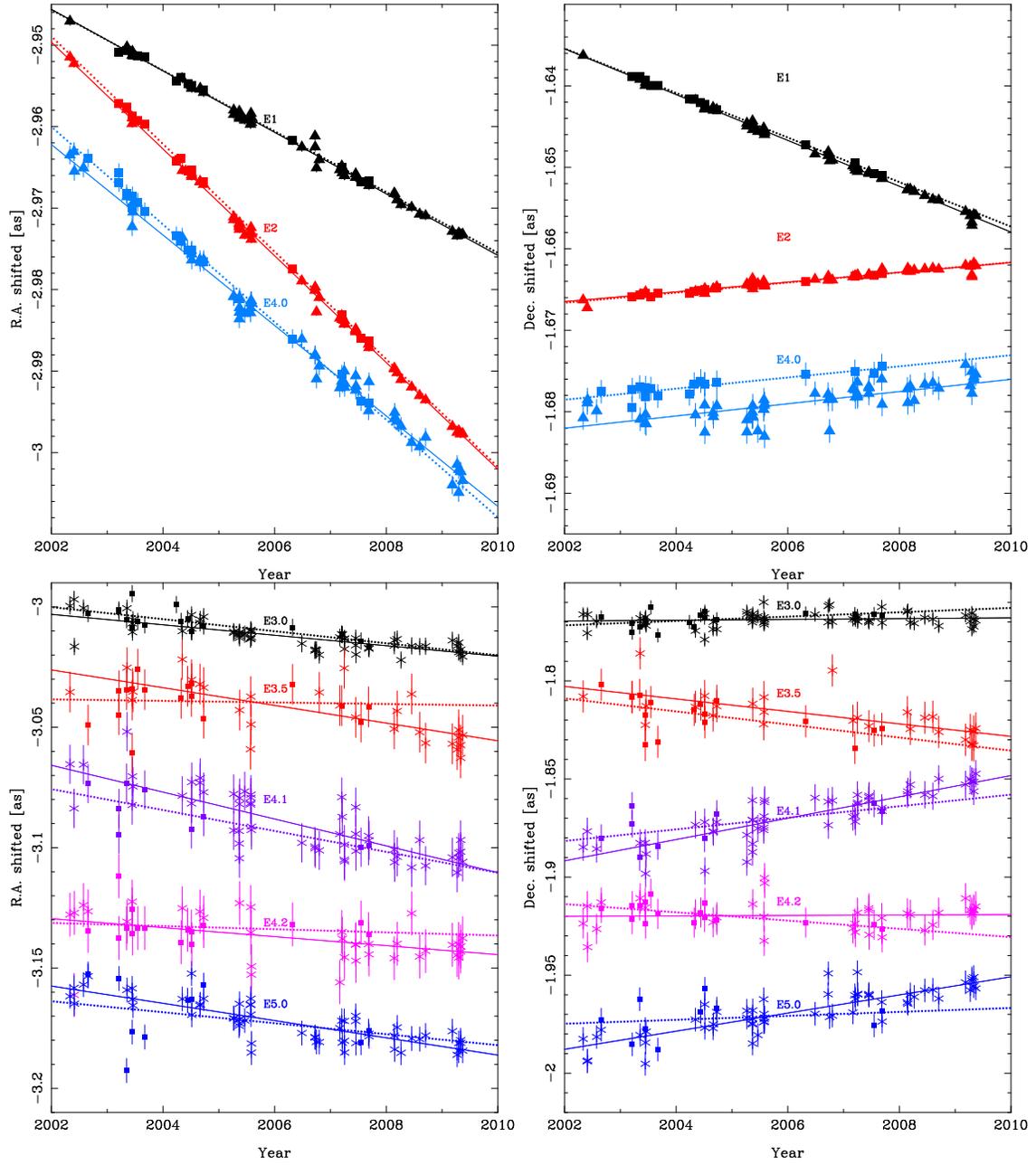

\begin{center}
\includegraphics[width=0.47\columnwidth,angle=-90]{fc1a.eps}
\includegraphics[width=0.47\columnwidth,angle=-90]{fc1b.eps}
\includegraphics[width=0.47\columnwidth,angle=-90]{fc1c.eps}
\includegraphics[width=0.47\columnwidth,angle=-90]{fc1d.eps}
\caption{Positions and linear motion fits for the objects in IRS13E. Top: The three bright stars in Ks-band (asterisks and solid lines) and H-band (boxes and dotted lines). Bottom: Some of the fainter objects. All data and fits are offset from each other for clarity. 
} 
\label{fig:positions1}
\end{center}
\end{figure}

\end{document}